\documentclass[12pt,useAMS,usenatbib,nomarkers,referee]{biom}

\usepackage[T1]{fontenc}
\usepackage[utf8]{inputenc}
\usepackage{lmodern}
\usepackage{textcomp}

\usepackage{fancyhdr}
\usepackage{amsmath,amssymb,bm,dsfont,mathrsfs}
\usepackage[mathscr]{euscript}
\allowdisplaybreaks

\usepackage{thmtools,thm-restate}

\usepackage{graphicx}
\graphicspath{{./figs/}}
\usepackage{xcolor}
\usepackage{float}
\usepackage{subcaption}   

\usepackage{setspace}
\usepackage{enumitem}
\usepackage{titlesec}
\titlespacing*{\section}{0pt}{.25\baselineskip}{.1\baselineskip}
\titlespacing*{\subsection}{0pt}{.25\baselineskip}{.1\baselineskip}
\titlespacing*{\subsubsection}{0pt}{0.25em}{0.25em}

\usepackage{booktabs}
\usepackage{multirow}
\usepackage{tabularx}
\usepackage{threeparttable}
\usepackage{pdflscape}

\makeatletter
\@ifundefined{thead}{}{}
\makeatother
\usepackage{makecell}

\usepackage{siunitx}
\newcolumntype{L}{>{\raggedright\arraybackslash}X}
\newcolumntype{C}{>{\centering\arraybackslash}m{0.9cm}} 
\newcolumntype{A}{S[table-format=1.3]} 
\newcolumntype{O}{S[table-format=1.3]} 
\newcolumntype{R}{S[table-format=1.2]} 

\newcommand{\na}{\textemdash}
\newcommand{\pblock}[1]{\multicolumn{3}{c}{#1}}

\usepackage{algorithm}
\usepackage{algpseudocode}

\usepackage{authblk}
\usepackage{multicol}
\usepackage{comment}
\usepackage[normalem]{ulem} 
\usepackage{url}

\usepackage[hidelinks]{hyperref}

\usepackage{orcidlink}

\title{Robust fuzzy clustering for high-dimensional multivariate time series with outlier detection}

  \author{Ziling Ma\,\orcidlink{0009-0006-0323-6467}\textsuperscript{1}, Ángel López-Oriona\,\orcidlink{0000-0003-1456-7342}\textsuperscript{1}, Hernando Ombao\,\orcidlink{0000-0001-7020-8091}\textsuperscript{1}, Ying Sun\,\orcidlink{0000-0001-6703-4270}\textsuperscript{1}
 \thanks{King Abdullah University of Science and Technology (KAUST), Computer, Electrical and Mathematical Sciences and Engineering (CEMSE)
 Division. Thuwal 23955-6900, Saudi Arabia. 
 Correspondence:
  ziling.ma@kaust.edu.sa, angel.lopezoriona@kaust.edu.sa, hernando.ombao@kaust.edu.sa, ying.sun@kaust.edu.sa} }

\begin{document}
\pagerange{\pageref{firstpage}--\pageref{lastpage}} \pubyear{2024}

\label{firstpage}


\begin{abstract}
\thispagestyle{empty}
Fuzzy clustering provides a natural framework for modeling partial memberships, particularly important in multivariate time series (MTS) where state boundaries are often ambiguous. For example, in EEG monitoring of driver alertness, neural activity evolves along a continuum (from unconscious to fully alert, with many intermediate levels of drowsiness) so crisp labels are unrealistic and partial memberships are essential. However, most existing algorithms are developed for static, low-dimensional data and struggle with temporal dependence, unequal sequence lengths, high dimensionality, and contamination by noise or artifacts. To address these challenges, we introduce RFCPCA, a robust fuzzy subspace-clustering method explicitly tailored to MTS that, to the best of our knowledge, is the first of its kind to simultaneously: (i) learn membership-informed subspaces, (ii) accommodate unequal lengths and moderately high dimensions, (iii) achieve robustness through trimming, exponential reweighting, and a dedicated noise cluster, and (iv) automatically select all required hyperparameters. These components enable RFCPCA to capture latent temporal structure, provide calibrated membership uncertainty, and flag series-level outliers while remaining stable under contamination. On driver drowsiness EEG, RFCPCA improves clustering accuracy over related methods and yields a more reliable characterization of uncertainty and outlier structure in MTS.

\end{abstract}


\maketitle
\clearpage
\pagenumbering{arabic}
\setcounter{page}{2}
\pagestyle{plain} 

\section{Introduction} \label{intro}

Time series data are ubiquitous across finance, medicine, engineering, and neuroscience, and modern acquisition technologies now generate massive repositories of recordings. These datasets are often heterogeneous in length, high-dimensional, and exhibit evolving temporal dynamics, making traditional mining tasks such as forecasting, classification, anomaly detection, and clustering challenging to scale and interpret \citep{esling2012time,ghassempour2014clustering}. The challenge is even greater for multivariate time series (MTS), where complex inter-channel dependencies must be preserved to meaningfully characterize system dynamics.

Clustering is a central tool for exploring and simplifying such data. By grouping MTS with similar temporal structures and cross-channel interactions \citep{aghabozorgi2015time}, clustering reduces data complexity, supports anomaly detection through deviations from cluster prototypes \citep{li2021clustering,shi2023clustering}, and facilitates predictive modeling tailored to homogeneous subgroups \citep{kavitha2010clustering,jha2015clustering}. However, in real-world MTS applications such as EEG or sensor data, clustering performance is often undermined by noise, artifacts, and outliers. High dimensionality further exacerbates the problem: distance measures lose discriminative power due to the curse of dimensionality, and even mild contamination can distort cluster assignments \citep{rousseeuw1991robust,aggarwal2001outlier,garcia2010review}. Although robust clustering methods exist \citep{dave2002robust}, they are typically computationally expensive and not well suited for large-scale or high-dimensional MTS.

Traditional clustering methods follow a hard (crisp) paradigm, where each element is assigned to exactly one group, whereas fuzzy clustering techniques offer a more flexible alternative. Many real datasets contain inherently ambiguous boundaries, making partial memberships more realistic \citep{hwang2007fuzzy,kruse2007fundamentals}. Beyond interpretability, fuzzy clustering also provides algorithmic advantages:
(i) smoother membership updates reduce sensitivity to poor initialization and local minima \citep{heiser1997cluster,jayaram2013can},
(ii) optimization tends to be faster and more stable \citep{mcbratney1985application,suganya2012fuzzy}, and
(iii) secondary memberships offer valuable diagnostic insights, highlighting borderline or anomalous samples that hard clustering ignores \citep{duran2013cluster,d2014robust,ruspini2019fuzzy}.
Despite these advantages, existing fuzzy clustering techniques rarely combine robustness, scalability, and the ability to handle high-dimensional MTS, leaving an important research gap.

In the previous literature on fuzzy clustering of MTS, \cite{d2012wavelets} introduced a wavelet-based method that groups series by similarities in their time–frequency patterns. \cite{izakian2015fuzzy} extended fuzzy clustering to MTS using dynamic time warping (DTW) distances, and \cite{d2021trimmed} later proposed a robust trimmed version of DTW-based clustering. \cite{li2020fuzzy} presented a fuzzy clustering approach with feature weights, emphasizing informative variables while down-weighting irrelevant ones. \cite{lopez2022quantile} proposed feature-based robust approaches for fuzzy clustering of MTS based on quantile cross-spectral density, enabling adaptive weighting of variables and distances. \cite{d2022spatial} developed a spatially robust fuzzy clustering framework for COVID-19 time series, where B-spline smoothing captures temporal trajectories and spatial dependence enhances clustering performance while reducing noise and outlier effects. \cite{d2023robust} proposed a fuzzy c-medoids method that integrates DTW with an exponential transformation and entropy regularization, improving robustness to outliers and accommodating unequal-length series. \cite{d2023owa} introduced a robust fuzzy clustering model for MTS based on a Student-t mixture, which down-weights the influence of outliers during clustering. Most recently, \cite{ma2025fuzzcoh} developed FuzzCoh, a robust spectral fuzzy clustering method using Kendall’s tau canonical coherence to capture frequency-domain connectivity, offering resilience to artifacts and strong interpretability for brain data. Nevertheless, existing fuzzy clustering methods for MTS remain computationally demanding, insufficiently robust to outliers, and often unable to handle unequal-length, high-dimensional series.

To address these challenges, we propose RFCPCA, a robust fuzzy clustering framework grounded in common principal component analysis (CPCA).
RFCPCA advances the state of the art in four key aspects:
\begin{enumerate}[leftmargin=*, itemsep=0.25em, topsep=0.25em]
\item It learns membership-weighted, cluster-specific low-rank subspaces via CPCA, linking soft memberships to the estimated bases while preserving cross-channel structure.
\item It naturally accommodates unequal-length and moderately high-dimensional series, avoiding restrictive preprocessing steps.
\item It integrates three complementary robustness strategies—exponential metric, trimming, and a noise cluster—that detect and mitigate outliers without sacrificing scalability.
\item It supports automatic hyperparameter selection, eliminating the need for ad hoc tuning.
\end{enumerate}

In summary, RFCPCA is, to our knowledge, the first subspace-based fuzzy clustering framework for MTS that jointly achieves robustness, interpretability, and computational feasibility.

The remainder of the paper is organized as follows. Section~\ref{methodology_intro} introduces the RFCPCA methodology. Section~\ref{numerical_results} evaluates its robustness on simulated contaminated EEG data. Section~\ref{application} investigates driver-drowsiness EEG data to assess how the learned structure aligns with levels of alertness and quantifies gains in accuracy, including principled series-level outlier handling. Finally, Section~\ref{conclusion} concludes the paper.

\section{RFCPCA: the robust fuzzy clustering method based on CPCA} \label{methodology_intro}
\cite{li2019multivariate} proposed Mc2PCA, a method for MTS clustering based on CPCA. This approach was subsequently improved by \cite{ma2024robcpca}, who introduced ROBCPCA, a robust version of Mc2PCA that additionally accounts for the serial dependence inherent in MTS. Building on the ROBCPCA framework, \cite{ma2025fcpca} developed FCPCA, a fuzzy clustering method grounded in CPCA. Their work demonstrated that FCPCA offers advantages in terms of computational efficiency over ROBCPCA, as well as improved clustering accuracy compared to both ROBCPCA and Mc2PCA. In what follows, we first provide an overview of the FCPCA method before presenting our proposed robust framework, RFCPCA, which enhances FCPCA by effectively handling outliers in MTS fuzzy clustering.

\subsection{Brief overview of FCPCA method} \label{FCPCA}

Consider a dataset $\bm{X}=\{\bm{X}_1,\dots,\bm{X}_N\}$ consisting of $N$ MTS, where each $\bm{X}_i \in \mathbb{R}^{T_i \times p}$ represents the $i$-th MTS object with length $T_i$ and dimensionality (channels) $p$. Each series is assumed second-order stationary: it has a constant mean and an autocovariance that depends only on the lag $\ell \in \mathbb{Z}$. In line with \cite{ma2025fcpca}, we restrict attention to nonnegative lags up to $L=2$ (i.e., $\ell \in \{0,1,2\}$).

The primary aim of FCPCA is to cluster the MTS dataset into $S$ fuzzy clusters, where each MTS object can have partial memberships across clusters. Initially, fuzzy membership degrees $u_{is}$ (for $i \in \{1,\dots,N\}$ and $s \in \{1,\dots,S\}$), representing the membership of the $i$-th object to cluster $s$, are randomly initialized, ensuring each object's memberships sum to one across all clusters, i.e., $\sum_{s=1}^{S} u_{is}=1$.

FCPCA first computes cross-covariance matrices at each lag for every series. For each MTS $\bm{X}_i$, the lagged cross-covariance matrices $\hat{\Gamma}_i(l)$ are estimated, and corresponding block covariance matrices are constructed:
$$\hat{\mathbf{\Gamma}}_i(l) = \begin{pmatrix}
\hat{\Gamma}_i(0) & \hat{\Gamma}_i(l) \\
\hat{\Gamma}_i(l)^\top & \hat{\Gamma}_i(0)
\end{pmatrix}, \quad l = 1, \dots,L.$$

Subsequently, common weighted covariance matrices for each cluster $s$ at lag $l$ are calculated using fuzzy memberships:
$$\tilde{\mathbf{\Sigma}}_s(l)=\frac{\sum_{i=1}^N u_{is}^m\hat{\mathbf{\Gamma}}_i(l)}{\sum_{i=1}^N u_{is}^m},\quad s=1,\dots,S,$$
where $m>1$ controls the fuzziness of the clustering, with a higher value of $m$ gives more fuzziness. 

Singular value decomposition (SVD) of each $\tilde{\mathbf{\Sigma}}_s(l)$ produces eigenvectors, from which the common projection axes $\mathbf{C}_s(l)$ are derived by selecting principal components that capture a substantial portion (e.g., 95\%) of the variance.

Each object $\hat{\mathbf{X}}_i(l)$ is formed by concatenating two lagged segments of the original MTS 
$\mathbf{X}_i$: the past segment $\mathbf{X}^*_{i,t-l}$ and the future segment $\mathbf{X}^*_{i,t}$. 
This construction ensures conformable dimensions for projection. The resulting 
$\hat{\mathbf{X}}_i(l)$ is then projected onto, and reconstructed from, the subspace of each cluster:
\[
\mathbf{Y}_i^{s}(l) = \hat{\mathbf{X}}_i(l) \, \mathbf{C}_s(l) \, \mathbf{C}_s(l)^\top, 
\quad s \in \{1,\dots,S\}.
\]
The reconstruction error, squared Frobenius distance, for each series in cluster $s$ across all lags is:
$$r_{is}^{2}=\sum_{l=1}^{L}\left\|\hat{\mathbf X}_i(l)-\bm{Y}_i^s(l)\right\|_{F}^{2}.$$
The primary objective of FCPCA is to find the $N\times S$ matrix of fuzzy membership, $\mathbf{U} = (u_{is})$, and sets of common projection axes, ${\bm{\mathcal{C}}}(l) = \{{\mathbf{C}}_1(l), \ldots, {\mathbf{C}}_S(l)\}$ that minimize the total weighted reconstruction error:
\begin{equation}\label{fuzzyCPCA_mini}
\left\{
\begin{aligned}
    &\min_{\mathbf{U}, {\bm{\mathcal{C}}}(l)} \quad \sum_{i=1}^{N}\sum_{s=1}^{S}u_{is}^m r_{is}^2 \\
    &\text{subject to} \quad \sum_{s=1}^{S} u_{is} = 1 \quad \text{and} \quad u_{is} \geq 0.
\end{aligned}
\right.
\end{equation}
The optimization problem is solved iteratively, and the membership degrees are updated based on reconstruction errors as:
$$u_{is}=\left[\sum_{s^*=1}^{S}\left(\frac{r_{is}^2}{r_{is^*}^2}\right)^{\frac{1}{m-1}}\right]^{-1}.$$
The algorithm alternates between updating fuzzy memberships and updating the common projection axes, and stops when the total reconstruction error converges (absolute change $<10^{-3}$) or when 1000 iterations are reached.

Upon convergence, we obtain the membership matrix $\mathbf{U}$, the common projection axes ${\bm{\mathcal{C}}}(l)$, and also the cluster prototype $\mathcal{P}\;=\;
\bigl\{\,P_{s}(l)\bigr\}, \: s \in \{1,\dots,S\},\: l \in \{1,\dots,L\}$, where $P_s(l) = {\mathbf{C}}_s(l)\,{\mathbf{C}}_s(l)^\top$.

\subsection{Robust metric approach: RFCPCA‑E} \label{RFCPCA_E}
In FCPCA, the squared reconstruction error determines the membership degree of each series, with larger errors reducing its weight in the assigned cluster. However, due to the quadratic loss, the influence of large errors remains unbounded \citep{rousseeuw1984least}. Even with slightly reduced membership, a large squared error resulted by an an outlier can dominate the objective and distort the common subspace. To address this, we replace the squared loss with an exponential loss:

\begin{equation}\label{eq:RFCPCA-E}
\left\{
\begin{array}{l@{\quad}l}
\displaystyle \min_{\mathbf{U},\,\bm{\mathcal C}(l)} &
\displaystyle \sum_{i=1}^{N}\sum_{s=1}^{S} u_{is}^{\,m}\!\left[1-\exp(-\beta r_{is}^{2})\right] \\[6pt]
\text{subject to} &
\displaystyle \sum_{s=1}^{S} u_{is}=1,\; u_{is}\ge 0,\; i=1,\ldots,N,
\end{array}
\right.
\end{equation}
where $\beta>0$ is a constant scale parameter controls how aggressively large reconstruction errors are downweighted in the exponential loss. Empirically, it can be selected as \citep{d2015time,lopez2022quantile}: 
\begin{equation}
  \beta=\left(\frac{1}{N}\sum_{i=1}^{N}\min_{1\le s\le S}r_{is}^{2}\right)^{-1}.
\end{equation}

The exponential distance is inherently more robust than the Euclidean distance, as it adaptively downweights outliers and gives greater weight to points near the bulk of the data \citep{wu2002alternative,lafuente2020robust}.

The local optimal membership update for \eqref{eq:RFCPCA-E} is given as follows:
\begin{equation}
    u_{is}
=\left[
\sum_{s'=1}^{S}
\left(
\frac{1-\exp\!\left(-\beta r_{is}^{2}\right)}
     {1-\exp\!\left(-\beta r_{is'}^{2}\right)}
\right)^{\!\frac{1}{m-1}}
\right]^{-1}.
\end{equation}

Since $x\mapsto 1-\exp(-\beta x)$ saturates at~1 as $x\to\infty$, outliers have bounded influence, unlike the unbounded penalty of the classical $\ell_{2}$ loss. When an MTS is highly outlying, the exponential loss yields nearly equal costs across clusters, resulting in evenly distributed memberships.


\subsection{Noise cluster approach: RFCPCA-N}\label{RFCPCA_N}

The noise cluster approach introduces an additional cluster dedicated to outliers. In this framework, the first $S-1$ clusters capture the genuine structure in the data, while cluster $S$ serves as a noise cluster characterized by a fixed noise distance $\delta > 0$. The parameter $\delta$ defines a dissimilarity threshold: objects whose minimum reconstruction error across all regular clusters exceeds $\delta$ are considered too distant and therefore receive higher membership in the noise cluster. Conceptually, $\delta$ delineates the boundary between informative observations and potential outliers \citep{cimino2004noise}. 

This mechanism enables the algorithm to assign atypical series to the noise cluster rather than forcing them into an inappropriate group, thereby preventing outliers from distorting the estimation of cluster-specific subspaces.

The optimization problem is formulated as
\begin{equation}\label{eq:RFCPCA-N}
\left\{
\begin{array}{ll}
\displaystyle 
\min_{\mathbf{U},\,{\bm{\mathcal{C}}}(l)} &
\displaystyle 
\sum_{i=1}^{N}\sum_{s=1}^{S-1}u_{is}^{\,m}r_{is}^{2}
+\sum_{i=1}^{N}\delta^{2}\!\left(1-\sum_{s=1}^{S-1}u_{is}\right)^{m} \\[10pt]
\text{subject to} &
\displaystyle 
\sum_{s=1}^{S}u_{is}=1,\quad u_{is}\ge0,\quad i=1,\dots,N.
\end{array}
\right.
\end{equation}

The membership update for the regular clusters is
\begin{equation}
  u_{is}=\Biggl[\sum_{s'=1}^{S-1}\Bigl(\tfrac{r_{is}^{2}}{r_{is'}^{2}}\Bigr)^{\!\frac{1}{m-1}}+\Bigl(\tfrac{r_{is}^{2}}{\delta^{2}}\Bigr)^{\!\frac{1}{m-1}}\Biggr]^{-1},\qquad s=1,\dots,S-1.
\end{equation}
The noise-cluster membership is then
\begin{equation}
  u_{iS}=1-\sum_{s=1}^{S-1}u_{is}.
\end{equation}

\begin{figure}[htbp]
    \centering
    \includegraphics[width=0.8\linewidth]{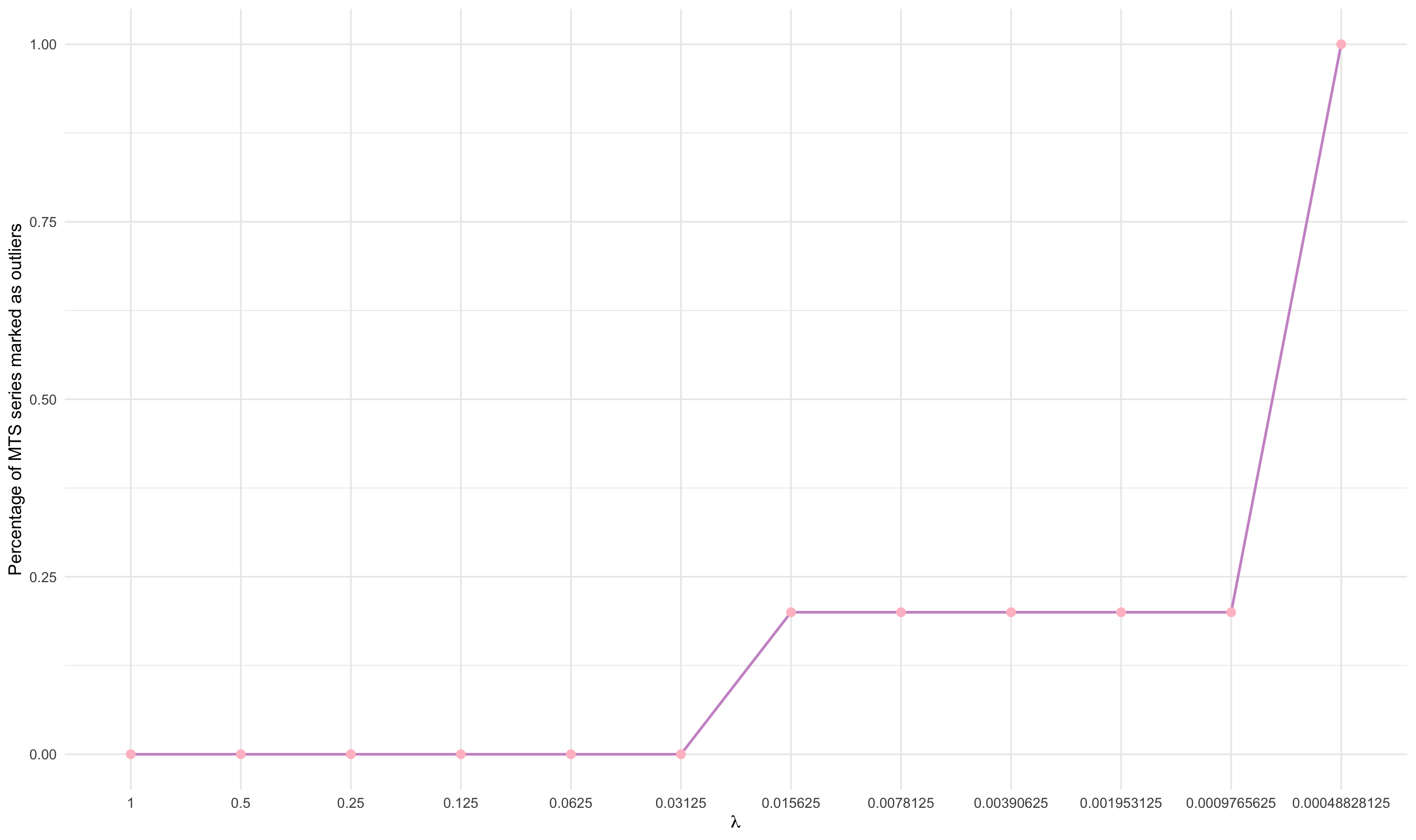}
  \caption{Example of outlier proportion versus $\lambda$, with the elbow point indicating the selected parameter.}
    \label{selection_lambda}
\end{figure}

Suggested by \cite{dave1991characterization}, the noise distance $\delta$ is updated as
\begin{equation}
  \delta^{2}=\frac{\lambda}{N\,(S-1)}\sum_{i=1}^{N}\sum_{s=1}^{S-1}r_{is}^{2},
\end{equation}
where $\lambda$ is a scale multiplier. Following \cite{dave1991characterization,rehm2007novel}, $\lambda$ can be selected by iteratively reducing its value and monitoring the proportion of objects assigned to the noise cluster. This proportion typically remains stable until $\lambda$ becomes so small that genuine cluster members start being absorbed into the noise cluster, producing a distinct “elbow” in the curve. The value of $\lambda$ at this elbow is taken as the optimal noise-distance threshold. However, as noted in \citet{dave1991characterization,d2021robust}, the results are generally not very sensitive to the choice of $\lambda$. Fig.~\ref{selection_lambda} shows an example of the selection of $\lambda$. Starting from $\lambda = 1$, no MTS objects are identified as outliers. As $\lambda$ decreases, the outlier proportion remains at 0\% until about $\lambda = 0.0156$, where it increases to roughly 20\% and then stabilizes. 
When $\lambda$ is further reduced to approximately $0.0004$, the proportion of outliers 
rises sharply to nearly 100\%. The optimal $\lambda$ is chosen at the elbow point, just before this abrupt increase, i.e., $\lambda \approx 0.0009$.

\subsection{Trimming approach: RFCPCA‑T} \label{RFCPCA_T}
The trimming approach is a classical robust strategy designed to mitigate the influence of outliers. Instead of incorporating all series into the estimation, a fixed proportion of the objects with the largest errors is excluded from contributing to the update of the cluster subspaces. This ensures that the estimated projection axes and memberships are driven by the bulk of the data, rather than being distorted by a small number of atypical series.  

Fix a trimming proportion $\alpha\in(0,1)$ and let $H=\lfloor N(1-\alpha)\rfloor$. Denote by $Y\subset\{1,\dots,N\}$ the index set of the $H$ retained series, so that only a subset of size $H$ of the $N$ available MTS objects is retained in the optimization. The objective function is then given by  
\begin{equation}\label{eq:RFCPCA-T}
\left\{
\begin{array}{ll}
\displaystyle 
\min_{\mathbf{U},\,{\bm{\mathcal{C}}}(l),\,Y:\,|Y|=H} &
\displaystyle 
\sum_{i\in Y}\sum_{s=1}^{S}u_{is}^{\,m}r_{is}^{2} \\[10pt]
\text{subject to} &
\displaystyle 
\sum_{s=1}^{S}u_{is}=1,\quad u_{is}\ge0,\quad i=1,\dots,N.
\end{array}
\right.
\end{equation}

The trimming set $Y$ is selected adaptively: at each iteration, the $H$ objects with the smallest reconstruction errors are retained, while the remaining $N-N$ series are excluded from updating the centroids and projection axes. In this way, the algorithm focuses on the bulk of the data while discarding potentially influential outliers.  

When $\alpha=0$, all series are retained and RFCPCA-T reduces to the standard FCPCA method. For $\alpha>0$, the effective sample size is reduced to $H$, which increases robustness at the expense of efficiency.

\subsection{Methodology comparison}

The three above methods are designed to address the sensitivity of FCPCA to outliers in MTS. Each method operates through a distinct mechanism and presents unique advantages and disadvantages.

RFCPCA-E  replaces the conventional $L_2$ norm with an exponential distance that progressively down-weights large errors \citep{banerjee2012robust}. This soft penalization reduces the influence of outliers while retaining all data, ensuring that every series contributes to clustering. It offers continuous and adaptive robustness without requiring trimming rates or extra clusters, thereby preserving model simplicity. However, extremely deviant outliers may still exert residual influence, explicit outlier labeling is not available, and performance depends on careful tuning of the scale parameter.

RFCPCA-N introduces an additional cluster dedicated to noise or outliers \citep{rehm2007novel}. All data are retained, but poorly fitting series are given higher membership in the noise cluster, allowing explicit identification of anomalies. This is particularly valuable for applications such as fault detection, neuroscience, finance, etc., where outlier behavior is informative. However, the approach increases model complexity and parameter tuning requirements, and borderline cases may receive ambiguous memberships, complicating interpretation.

RFCPCA-T excludes a fixed proportion of the most outlying series based on reconstruction errors \citep{garcia2008general,garcia2024robust}. By removing these points entirely, the estimation of cluster parameters becomes more stable and resistant to extreme contamination. Trimming thus provides strong robustness and computational efficiency through reduced sample size, with clear separation between retained and discarded data. The drawback is that trimmed series are permanently excluded, which risks discarding meaningful signals, and the method does not adapt to varying levels of outlyingness.

In summary, trimming is most effective when extreme outliers must be eliminated, the noise cluster is preferable when retaining and labeling anomalies is important, and the robust metric is best suited when continuous robustness is desired without discarding data or modifying the clustering structure.

\subsection{Selection of hyperparameters}
To run the RFCPCA algorithm, the number of clusters $S$ and the fuzziness parameter $m$ must be specified (and, for the trimmed variant RFCPCA-T, the trimming proportion $\alpha$). 
Because different hyperparameter choices can yield markedly different partitions and are rarely known in advance \citep{arbelaitz2013extensive}, a validation criterion is required to identify the optimal clustering \citep{kwon1998cluster,wang2007fuzzy,wang2022survey}.

Following \cite{ma2025fcpca}, we consider the generalization of the Xie-Beni index \citep{85677} as it directly uses the total weighted reconstruction error that FCPCA and RFCPCA present when the computation ends.

Denote by $\varphi$ the loss function corresponding to the clustering method under evaluation (FCPCA, RFCPCA-E, RFCPCA-N, or RFCPCA-T).  
We define
\begin{equation}
    J(\mathbf{U},\bm{\mathcal{C}}(l);\varphi) =
    \sum_{i=1}^N \sum_{s=1}^S u_{is}^m \, \varphi\bigl(r_{is}^2\bigr),
\end{equation}
to provide a unified formulation and avoid redundancy.

Let
\begin{equation}
    d_{\min} = \min_{s \neq s'} \sum_{l=1}^{L}
      \bigl\lVert \mathbf{P}_{s}(l) - \mathbf{P}_{s'}(l) \bigr\rVert_{F}^{2},
      \quad s,s' \in \{1,\dots,S\},
\end{equation}
denote the minimum distance between the subspace prototypes of any two distinct clusters.

We then define the cluster validity index (CVI) as
\begin{equation}
     \operatorname{CVI}(\varphi)
  =\frac{J(\mathbf{U},\bm{\mathcal{C}}(l);\varphi)}
         {N \, d_{\min}}.
\end{equation}

For every candidate tuple $(S,m)$ (or triplet $(S,m,\alpha)$ in the case of RFCPCA-T), we run the algorithm until convergence and record $\operatorname{CVI}(\varphi)$. The numerator measures the within-cluster reconstruction loss under the given method, while the denominator quantifies the separation between clusters at the end of the clustering process. Hence, lower values of $\operatorname{CVI}(\varphi)$ are preferred.

\section{Numerical results}\label{numerical_results}
We focus on the simulated contaminated EEG data to evaluate the robustness of RFCPCA. For completeness and reproducibility, we first specify the clean EEG generator and then the artifact models. Our method is also feasible 

\subsection{Clean EEG generator (oscillatory mixing)}
Let $N=20$ trials (10 per group), each an $n\times p$ matrix with sampling rate $f_s=100$\,Hz. For each trial we generate five latent AR(2) processes corresponding to five frequency bands $\{\delta,\theta,\alpha,\beta,\gamma\}$. For band $b$ with peak frequency $f_b$ and sharpness $\kappa_b$, the AR(2) coefficients are
\[
\phi_1^{(b)}=\frac{2}{M_b}\cos\!\left(2\pi \frac{f_b}{f_s}\right),\qquad
\phi_2^{(b)}=-\frac{1}{M_b^2},\qquad M_b=\exp(\kappa_b),
\]
with $(f_\delta,f_\theta,f_\alpha,f_\beta,f_\gamma)=(2,6,10,22.5,37.5)$\,Hz and $(\kappa_\delta,\kappa_\theta,\kappa_\alpha,\kappa_\beta,\kappa_\gamma)=(0.05,0.05,0.05,0.08,0.10)$.
Each latent is standardized to unit variance and lightly band-pass filtered to its band.
Let $Z(t)\in\mathbb{R}^{1\times 5}$ stack the five latents at time $t$ and $A\in\mathbb{R}^{5\times p}$ be a nonnegative mixing matrix whose columns sum to one.
Group~1 channels are dominated by $\{\delta,\theta,\alpha,\gamma\}$. Group~2 channels are dominated by $\{\theta,\beta\}$. Thus, the two groups are inherently fuzzy.
Thus each trial is formed as
\[
X(t)=Z(t)\,A\in\mathbb{R}^{1\times p},\qquad t=1,\dots,T .
\]

\subsection{Comparison methods and evaluation metrics}
In this section, we present the baseline methods used for comparison and describe the evaluation metrics employed to assess clustering performance in the simulation studies.
\subsubsection{Alternative methods}
 
We compare RFCPCA with three widely used baselines for multivariate MTS as follows and we also evaluate their robust variants. 

\begin{itemize}[leftmargin=*, itemsep=2pt, topsep=2pt, parsep=0pt, partopsep=0pt]
    \item \textit{{WWW by \cite{wang2007structure}.}} WWW summarizes each univariate channel with a fixed-length vector of structure-based statistical features. The per-channel feature vectors are concatenated across channels and then clustered with standard fuzzy $C$-means.
    
\item \textit{QCD by \cite{lopez2021quantile}}. Each MTS is summarized by quantile cross-spectral density estimates over a grid of frequencies and quantile levels. The resulting complex-valued summaries are split into real and imaginary parts and concatenated into a single feature vector per series. Clustering is then performed with fuzzy $C$-means in this feature space—often after a PCA step for decorrelation/denoising—using Euclidean distance to centroids. This representation captures serial and cross-component dependence and is robust to heavy tails and extremes.

    \item \textit{MODWT by \cite{d2012wavelets}}. MODWT decomposes each channel via the MODWT and uses scale-specific wavelet variances (per channel) and wavelet correlations (across channel pairs) as features. The concatenated feature vector summarizes variability and inter-channel interaction across scales, and can be clustered with fuzzy $C$-means. MODWT retains time alignment, works for mean-nonstationary series, and makes inter-component relationships explicit.
\end{itemize}

The hyperparameters, namely $\beta$ and $\delta$, in the alternative methods are selected analogously to those in RFCPCA.

\subsubsection{Evaluation metrics}
For each scenario, we report (i) clustering accuracy on the non–outlier series (Acc) and (ii) the percentage of true outliers correctly detected (Out\%). For RFCPCA–T we also report the average trimming ratio selected by the cluster validity index (CVI).

To compare with ground-truth hard labels, fuzzy memberships are converted to hard labels using the threshold $0.70$ \citep{maharaj2011fuzzy,lopez2021quantile,ma2025fcpca}. For $S=2$,
\[
\text{label}_i \;=\; \arg\max_{k\in\{1,2\}} u_{ik}
\quad\text{if}\quad \max\{u_{i1},u_{i2}\}\ge 0.70.
\]

For each method, we present the outlier handling rules as follows.  
\begin{itemize}[leftmargin=*, itemsep=0.25em, topsep=0.25em, parsep=0pt]
\item \emph{Exponential metric (E).} An observation is flagged as an outlier if it has no dominant membership across the two substantive clusters, i.e., $\max\{u_{i1},u_{i2}\}<0.70$. If this occurs for a non-outlier series, it is counted as a false positive for outlier detection.

\item \emph{Noise cluster (N).} An observation is flagged as an outlier if its noise membership exceeds $0.50$, i.e., $u_{i,S}\ge 0.50$. For observations not flagged as outliers, memberships are renormalized over the non-noise clusters,
\[
\tilde u_{is} \;=\; \frac{u_{is}}{\sum_{s'\neq \text{S}} u_{is'}},
\]
and the $0.70$ threshold is then applied to $\{\tilde u_{ik}\}$. If no cluster attains the 0.70 threshold, the observation is counted as misclassified.

\item \emph{Trimming (T).} An observation is flagged as an outlier if it is in the trimmed set selected by the cluster validity index (CVI). Clustering accuracy is then computed on the remaining (non-trimmed) series.
\end{itemize}

\noindent\textbf{Metrics.}
Let $\mathcal{O}$ be the set of injected (true) outliers and $\widehat{\mathcal{O}}$ the set flagged by a given method. We report
\[
\text{Out\%} \;=\; 100\times \frac{|\widehat{\mathcal{O}}\cap \mathcal{O}|}{|\mathcal{O}|},
\]
i.e., the true-positive rate (recall) for outlier detection. Clustering accuracy is computed on the remaining series $\mathcal{I}=\{1,\ldots,N\}\setminus \widehat{\mathcal{O}}$ by the Rand index (RI) between ground-truth labels $\{y_i\}_{i\in \mathcal{I}}$ and the hard labels obtained as above, and reported as Acc.

\subsection{Simulated EEG data with transient electromyographic bursts} \label{burst_contamination_sim}
We evaluate on datasets with trial-level contamination. In awake EEG, scalp electromyographic (EMG) activity is a common artifact: it appears as brief, burst-like events with broadband spectra overlapping the \(\beta/\gamma\) range (overall \(\sim 20\text{--}300\,\mathrm{Hz}\), with substantial power in \(20\text{--}80\,\mathrm{Hz}\)). These bursts are typically spatially localized (strongest at peripheral electrodes) yet can project more broadly across the montage; their amplitudes can exceed neural high-frequency EEG by orders of magnitude, yielding heavy-tailed residuals under Gaussian error models \citep{goncharova2003emg,whitham2007scalp}. This setting stresses clustering methods and is well suited for assessing robustness.

In each group, we contaminate a fixed proportion \(\rho=0.20\) of trials at the trial level: specifically, $k_{\mathrm{pg}}=\left\lceil \rho\, N_{\mathrm{per\,group}}\right\rceil$ trials are selected uniformly at random without replacement and modified. All remaining trials stay clean.  

For each contaminated trial, draw \(n_{\mathrm{burst}}\in\{1,2,3\}\). For each burst, sample:
a start time \(t_0\in\{1,\dots,T_i-\tau\}\), a duration \(\tau=\lfloor 0.25\,f_s\rfloor\) (250\,ms at sampling rate \(f_s\)), a center frequency \(f^{\star}\sim\mathcal{U}(30,80)\,\mathrm{Hz}\), and a channel set \(\mathcal{I}\subset\{1,\dots,p\}\) with \(|\mathcal{I}|=\lceil 0.10\,p\rceil\).
With the Hann envelope \(h_q=\tfrac{1}{2}\bigl(1-\cos(2\pi q/(\tau-1))\bigr)\) for \(q=0,\dots,\tau-1\), define
\[
g_q=\sin\!\Bigl(2\pi f^{\star}\frac{q}{f_s}\Bigr)\,h_q,\qquad q=0,\dots,\tau-1.
\]
Let \(\hat{\sigma}_j\) denote the empirical standard deviation of channel \(j\) in that trial and set an amplitude multiplier \(\eta=5\). Update, for all \(q=0,\dots,\tau-1\) and \(j\in\mathcal{I}\),
\[
X_{t_0+q,\,j}\ \leftarrow\ X_{t_0+q,\,j}\ +\ \eta\,\hat{\sigma}_j\, g_q.
\]
Trials not selected remain clean.

\begin{figure}[ht]
    \centering
    \includegraphics[width=1\linewidth]{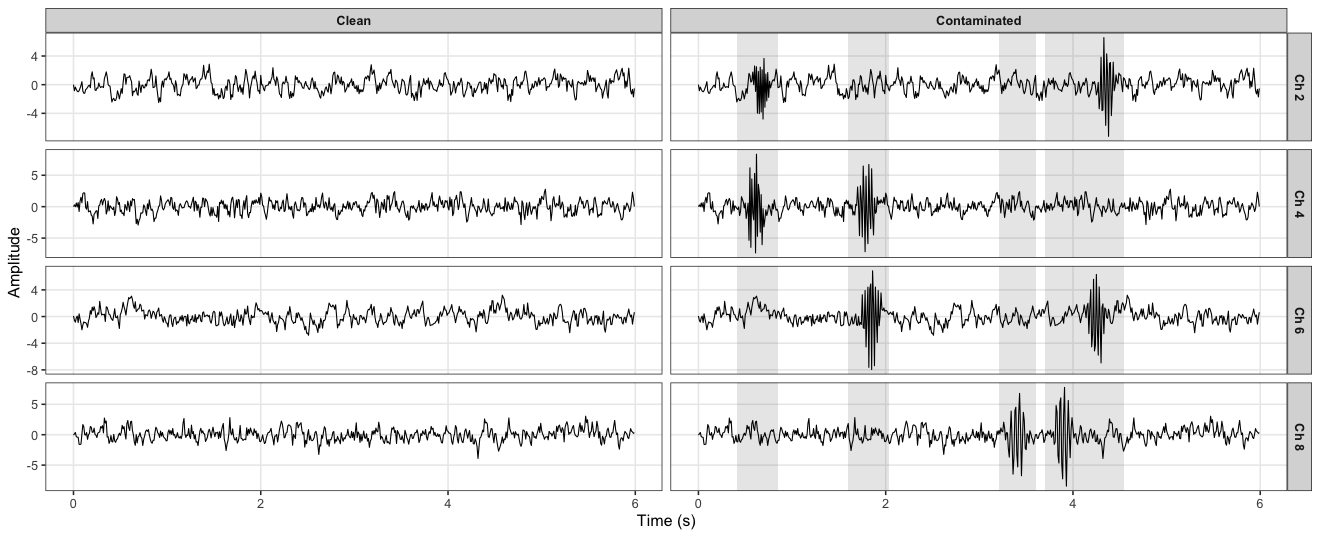}
    \caption{Simulated EEG data with burst contamination}
    \label{burst}
\end{figure}

\begin{figure}[ht]
    \centering
    \includegraphics[width=1\linewidth]{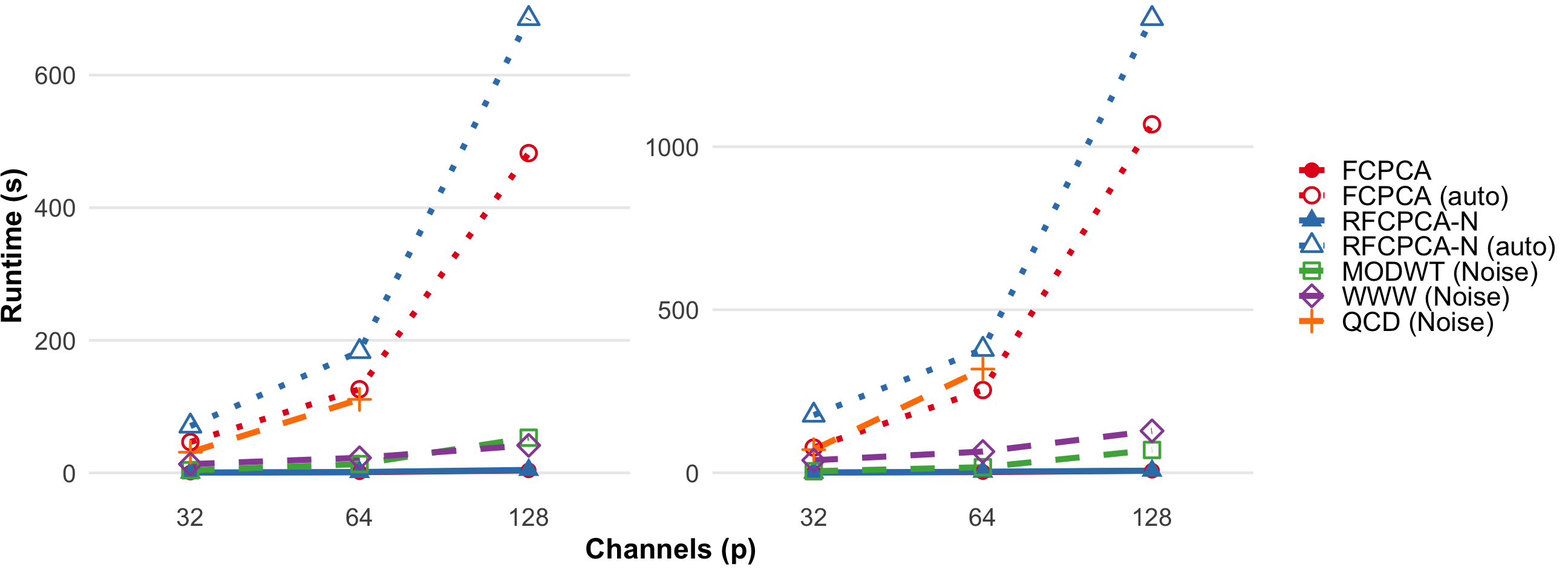}
    \caption{The run time comparison, the noise approach is used to represent the runtime of the robust methods. The left one is when $T=400$, right one is when $T=1000$.}
    \label{run_time}
\end{figure}

Fig.~\ref{burst} illustrates the simulated EEG data before and after the introduction of burst contamination. In total, 50 replications are used to obtain the results. Fig.~\ref{run_time} reports the average runtime of each method, where the noise-based variant is used as a reference since robust approaches generally require more computation. The runtime of QCD at $p=128$ is omitted due to memory limitations on the laptop. All computations are performed in R on a MacBook Pro (Apple M1 Max, 32 GB RAM).

From Fig.~\ref{run_time}, we observe that when $m$ and $k$ are fixed, both FCPCA and RFCPCA-N are among the fastest methods, and employing the robust variant does not incur a substantial additional cost. The dashed blue and orange lines show the runtime when $m$ is selected automatically from a grid $m \in \{1.1, 1.2, 1.4, 1.6, 1.8, 2.0, 2.2, 2.5\}$, which increases the runtime but remains within a feasible range. Notably, the QCD method requires runtimes comparable to or even longer than those of FCPCA with a grid search over $m$, underscoring that the proposed RFCPCA methods are computationally efficient and scalable, even when automatic parameter selection is incorporated.

Table~\ref{tab:auto_fcpca_rfcpca_tall} reports the results for FCPCA and the RFCPCA family with automatic selection of $m$ and $\alpha$ for RFCPCA-T suggested by the CVI. For comparison, Tables~\ref{T_400_com} and \ref{T_1000_com} summarize the results for the alternative methods for which we did not implement automatic tuning. Performance is reported over manually specified grids of $m$ and $\alpha$. Across settings with a contamination rate of 20\%, all methods achieve relatively high clustering accuracy.

RFCPCA delivers strong outlier handling without sacrificing accuracy. In particular, RFCPCA-T attains near-perfect detection (Out\% $\approx 0.95$–$1.00$) with Acc $=1.00$ in all configurations, while RFCPCA-E also maintains Acc $=1.00$ and detects a large fraction of bursts (Out\% $\approx 0.67$–$0.83$). The noise-cluster variant RFCPCA-N provides explicit anomaly labels and competitive detection (Out\% $\approx 0.68$–$0.78$), with a modest accuracy trade-off at larger $p$ due to membership mass assigned to the noise component. By contrast, FCPCA attains high accuracy but detects virtually no outliers. Even with manual tuning, the comparison methods in Tables~\ref{T_400_com}–\ref{T_1000_com} detect far fewer outliers (typically 0–6\% for E/N and 37–57\% for trimmed variants with fixed $\alpha$); notably, even a large fixed trimming level such as $\alpha=0.40$ does not remove all contaminated trials. Overall, the results highlight the advantage of automatic tuning in RFCPCA for accurate clustering and effective outlier detection while eliminating manual parameter search.

\begin{table}[ht]
\centering
\caption{FCPCA and RFCPCA with automatic selection of $m$ (and $\alpha$ for trimming).}
\label{tab:auto_fcpca_rfcpca_tall}
\setlength{\tabcolsep}{6pt}
\renewcommand{\arraystretch}{1.08}
\begin{tabular}{l c A O @{\,\,\,} R  A O @{\,\,\,} R  A O @{\,\,\,} R}
\toprule
\multirow{2}{*}{Method} & \multirow{2}{*}{$T$}
 & \pblock{$p=32$} & \pblock{$p=64$} & \pblock{$p=128$} \\
\cmidrule(lr){3-5}\cmidrule(lr){6-8}\cmidrule(lr){9-11}
 &  & Acc & Out\% & $\alpha$ & Acc & Out\% & $\alpha$ & Acc & Out\% & $\alpha$ \\
\midrule
\multirow{2}{*}{FCPCA}
 & 400  & 1.00 & 0.00 & \na & 1.00 & 0.00 & \na & 1.00 & 0.00 & \na \\
 & 1000 & 1.00 & 0.00 & \na & 1.00 & 0.00 & \na & 1.00 & 0.00 & \na \\
\addlinespace
\multirow{2}{*}{RFCPCA-E}
 & 400  & 1.00 & 0.75 & \na & 1.00 & 0.80 & \na & 1.00 & 0.83 & \na \\
 & 1000 & 1.00 & 0.68 & \na & 1.00 & 0.67 & \na & 1.00 & 0.75 & \na \\
\addlinespace
\multirow{2}{*}{RFCPCA-N}
 & 400  & 1.00 & 0.75 & \na & 0.98 & 0.78 & \na & 0.97 & 0.73 & \na \\
 & 1000 & 1.00 & 0.75 & \na & 0.95 & 0.68 & \na & 0.96 & 0.70 & \na \\
\addlinespace
\multirow{2}{*}{RFCPCA-T}
 & 400  & 1.00 & 1.00 & 0.50 & 1.00 & 0.98 & 0.50 & 1.00 & 0.95 & 0.50 \\
 & 1000 & 1.00 & 1.00 & 0.50 & 1.00 & 0.95 & 0.45 & 1.00 & 0.97 & 0.47 \\
\bottomrule
\end{tabular}
\end{table}

\begin{table}[htbp]
\centering
\small
\caption{Comparison methods results for $T=400$. Acc = accuracy, Out\% = outlier fraction.}
\label{T_400_com}
\begin{tabular*}{\linewidth}{@{\extracolsep{\fill}} ll *{3}{c@{\hspace{1.0em}}c} @{}}
\toprule
\multicolumn{2}{c}{} &
\multicolumn{2}{c}{$p=32$} &
\multicolumn{2}{c}{$p=64$} &
\multicolumn{2}{c}{$p=128$} \\
\cmidrule(lr){3-4}\cmidrule(lr){5-6}\cmidrule(lr){7-8}
Method & $m$
& Acc & Out\%
& Acc & Out\%
& Acc & Out\% \\
\midrule
\multirow{4}{*}{\textbf{WWW-E}}
& 1.5 & 1.00 & 0.05 & 1.00 & 0.04 & 1.00 & 0.04 \\
& 1.8 & 1.00 & 0.06 & 1.00 & 0.04 & 1.00 & 0.05 \\
& 2.0 & 1.00 & 0.06 & 1.00 & 0.05 & 1.00 & 0.05 \\
& 2.2 & 1.00 & 0.06 & 1.00 & 0.05 & 1.00 & 0.06 \\
\midrule
\multirow{4}{*}{\textbf{WWW-N}}
& 1.5 & 1.00 & 0.00 & 1.00 & 0.00 & 1.00 & 0.00 \\
& 1.8 & 1.00 & 0.00 & 0.77 & 0.24 & 1.00 & 0.00 \\
& 2.0 & 1.00 & 0.00 & 0.79 & 0.24 & 1.00 & 0.00 \\
& 2.2 & 1.00 & 0.00 & 0.85 & 0.24 & 1.00 & 0.00 \\
\midrule
\multicolumn{8}{l}{\textit{Trimmed (T) with } $\alpha=0.20$} \\
\multirow{4}{*}{\textbf{WWW-T}}
& 1.5 & 1.00 & 0.37 & 1.00 & 0.39 & 1.00 & 0.41 \\
& 1.8 & 1.00 & 0.37 & 1.00 & 0.39 & 1.00 & 0.41 \\
& 2.0 & 1.00 & 0.37 & 1.00 & 0.39 & 1.00 & 0.41 \\
& 2.2 & 1.00 & 0.37 & 1.00 & 0.39 & 1.00 & 0.41 \\
\midrule
\multicolumn{8}{l}{\textit{Trimmed (T) with } $\alpha=0.40$} \\
\multirow{4}{*}{\textbf{WWW-T}}
& 1.5 & 1.00 & 0.51 & 1.00 & 0.56 & 1.00 & 0.54 \\
& 1.8 & 1.00 & 0.51 & 1.00 & 0.56 & 1.00 & 0.52 \\
& 2.0 & 1.00 & 0.51 & 1.00 & 0.56 & 1.00 & 0.50 \\
& 2.2 & 1.00 & 0.51 & 1.00 & 0.57 & 1.00 & 0.49 \\
\midrule
\multirow{4}{*}{\textbf{MODWT-E}}
& 1.5 & 1.00 & 0.05 & 1.00 & 0.04 & 1.00 & 0.04 \\
& 1.8 & 1.00 & 0.06 & 1.00 & 0.05 & 1.00 & 0.05 \\
& 2.0 & 1.00 & 0.06 & 1.00 & 0.05 & 1.00 & 0.05 \\
& 2.2 & 1.00 & 0.07 & 1.00 & 0.07 & 1.00 & 0.06 \\
\midrule
\multirow{4}{*}{\textbf{MODWT-N}}
& 1.5 & 1.00 & 0.00 & 1.00 & 0.00 & 1.00 & 0.00 \\
& 1.8 & 1.00 & 0.00 & 0.76 & 0.25 & 1.00 & 0.00 \\
& 2.0 & 1.00 & 0.00 & 0.78 & 0.25 & 1.00 & 0.00 \\
& 2.2 & 1.00 & 0.00 & 0.81 & 0.25 & 1.00 & 0.00 \\
\midrule
\multicolumn{8}{l}{\textit{Trimmed (T) with } $\alpha=0.20$} \\
\multirow{4}{*}{\textbf{MODWT-T}}
& 1.5 & 1.00 & 0.39 & 1.00 & 0.37 & 1.00 & 0.39 \\
& 1.8 & 1.00 & 0.39 & 1.00 & 0.37 & 1.00 & 0.39 \\
& 2.0 & 1.00 & 0.39 & 1.00 & 0.37 & 1.00 & 0.39 \\
& 2.2 & 1.00 & 0.39 & 1.00 & 0.37 & 1.00 & 0.39 \\
\midrule
\multicolumn{8}{l}{\textit{Trimmed (T) with } $\alpha=0.40$} \\
\multirow{4}{*}{\textbf{MODWT-T}}
& 1.5 & 1.00 & 0.55 & 1.00 & 0.58 & 1.00 & 0.52 \\
& 1.8 & 1.00 & 0.55 & 1.00 & 0.58 & 1.00 & 0.53 \\
& 2.0 & 1.00 & 0.55 & 1.00 & 0.58 & 1.00 & 0.54 \\
& 2.2 & 1.00 & 0.55 & 1.00 & 0.57 & 1.00 & 0.54 \\
\midrule
\multirow{4}{*}{\textbf{QCD-E}}
& 1.5 & 1.00 & 0.05 & 1.00 & 0.04 & -- & -- \\
& 1.8 & 1.00 & 0.06 & 1.00 & 0.04 & -- & -- \\
& 2.0 & 1.00 & 0.06 & 1.00 & 0.05 & -- & -- \\
& 2.2 & 1.00 & 0.06 & 1.00 & 0.05 & -- & -- \\
\midrule
\multirow{4}{*}{\textbf{QCD-N}}
& 1.5& 1.00 & 0.00 & 1.00 & 0.00 & -- & -- \\
& 1.8 & 1.00 & 0.00 & 0.79 & 0.23 & -- & -- \\
& 2.0 & 1.00 & 0.00 & 0.79 & 0.24 & -- & -- \\
& 2.2 & 1.00 & 0.00 & 0.84 & 0.24 & -- & -- \\
\midrule
\multicolumn{8}{l}{\textit{Trimmed (T) with } $\alpha=0.20$} \\
\multirow{4}{*}{\textbf{QCD-T}}
& 1.5 & 1.00 & 0.37 & 1.00 & 0.39 & -- & -- \\
& 1.8 & 1.00 & 0.37 & 1.00 & 0.38 & -- & -- \\
& 2.0 & 1.00 & 0.36 & 1.00 & 0.38 & -- & -- \\
& 2.2 & 1.00 & 0.37 & 1.00 & 0.39 & -- & -- \\
\midrule
\multicolumn{8}{l}{\textit{Trimmed (T) with } $\alpha=0.40$} \\
\multirow{4}{*}{\textbf{QCD-T}}
& 1.5 & 1.00 & 0.51 & 1.00 & 0.56 & -- & -- \\
& 1.8 & 1.00 & 0.51 & 1.00 & 0.55 & -- & -- \\
& 2.0 & 1.00 & 0.52 & 1.00 & 0.53 & -- & -- \\
& 2.2 & 1.00 & 0.50 & 1.00 & 0.54 & -- & -- \\
\bottomrule
\end{tabular*}
\end{table}

\begin{table}[htbp]
\centering
\small
\caption{Comparison methods  results for $T=1000$. Acc = accuracy, Out\% = outlier fraction.}
\label{T_1000_com}
\begin{tabular*}{\linewidth}{@{\extracolsep{\fill}} ll *{3}{c@{\hspace{1.0em}}c} @{}}
\toprule
\multicolumn{2}{c}{} &
\multicolumn{2}{c}{$p=32$} &
\multicolumn{2}{c}{$p=64$} &
\multicolumn{2}{c}{$p=128$} \\
\cmidrule(lr){3-4}\cmidrule(lr){5-6}\cmidrule(lr){7-8}
Method & $m$
& Acc & Out\%
& Acc & Out\%
& Acc & Out\% \\
\midrule
\multirow{4}{*}{\textbf{WWW-E}}
& 1.5 & 1.00 & 0.03 & 1.00 & 0.04 & 1.00 & 0.04 \\
& 1.8 & 1.00 & 0.04 & 1.00 & 0.05 & 1.00 & 0.05 \\
& 2.0 & 1.00 & 0.04 & 1.00 & 0.05 & 1.00 & 0.05 \\
& 2.2 & 1.00 & 0.05 & 1.00 & 0.05 & 1.00 & 0.05 \\
\midrule
\multirow{4}{*}{\textbf{WWW-N}}
& 1.5 & 1.00 & 0.00 & 1.00 & 0.00 & 1.00 & 0.00 \\
& 1.8 & 1.00 & 0.00 & 0.74 & 0.24 & 1.00 & 0.00 \\
& 2.0 & 1.00 & 0.00 & 0.78 & 0.24 & 1.00 & 0.00 \\
& 2.2 & 1.00 & 0.00 & 0.83 & 0.24 & 1.00 & 0.00 \\
\midrule
\multicolumn{8}{l}{\textit{Trimmed (T) with } $\alpha=0.20$} \\
\multirow{4}{*}{\textbf{WWW-T}}
& 1.5 & 1.00 & 0.36 & 1.00 & 0.40 & 1.00 & 0.39 \\
& 1.8 & 1.00 & 0.36 & 1.00 & 0.40 & 1.00 & 0.39 \\
& 2.0 & 1.00 & 0.36 & 1.00 & 0.40 & 1.00 & 0.39 \\
& 2.2 & 1.00 & 0.36 & 1.00 & 0.40 & 1.00 & 0.39 \\
\midrule
\multicolumn{8}{l}{\textit{Trimmed (T) with } $\alpha=0.40$} \\
\multirow{4}{*}{\textbf{WWW-T}}
& 1.5 & 1.00 & 0.54 & 1.00 & 0.56 & 1.00 & 0.54 \\
& 1.8 & 1.00 & 0.54 & 1.00 & 0.56 & 1.00 & 0.54 \\
& 2.0 & 1.00 & 0.54 & 1.00 & 0.56 & 1.00 & 0.54 \\
& 2.2 & 1.00 & 0.54 & 1.00 & 0.57 & 1.00 & 0.54 \\
\midrule
\multirow{4}{*}{\textbf{MODWT-E}}
& 1.5 & 1.00 & 0.03 & 1.00 & 0.04 & 1.00 & 0.04 \\
& 1.8 & 1.00 & 0.04 & 1.00 & 0.05 & 1.00 & 0.06 \\
& 2.0 & 1.00 & 0.04 & 1.00 & 0.05 & 1.00 & 0.06 \\
& 2.2 & 1.00 & 0.06 & 1.00 & 0.05 & 1.00 & 0.06 \\
\midrule
\multirow{4}{*}{\textbf{MODWT-N}}
& 1.5 & 1.00 & 0.00 & 1.00 & 0.00 & 1.00 & 0.00 \\
& 1.8 & 1.00 & 0.00 & 0.76 & 0.25 & 1.00 & 0.00 \\
& 2.0 & 1.00 & 0.00 & 0.78 & 0.25 & 1.00 & 0.00 \\
& 2.2 & 1.00 & 0.00 & 0.81 & 0.25 & 1.00 & 0.00 \\
\midrule
\multicolumn{8}{l}{\textit{Trimmed (T) with } $\alpha=0.20$} \\
\multirow{4}{*}{\textbf{MODWT-T}}
& 1.5 & 1.00 & 0.37 & 1.00 & 0.40 & 1.00 & 0.39 \\
& 1.8 & 1.00 & 0.37 & 1.00 & 0.40 & 1.00 & 0.39 \\
& 2.0 & 1.00 & 0.37 & 1.00 & 0.40 & 1.00 & 0.39 \\
& 2.2 & 1.00 & 0.37 & 1.00 & 0.40 & 1.00 & 0.39 \\
\midrule
\multicolumn{8}{l}{\textit{Trimmed (T) with } $\alpha=0.40$} \\
\multirow{4}{*}{\textbf{MODWT-T}}
& 1.5 & 1.00 & 0.55 & 1.00 & 0.58 & 1.00 & 0.53 \\
& 1.8 & 1.00 & 0.55 & 1.00 & 0.58 & 1.00 & 0.53 \\
& 2.0 & 1.00 & 0.55 & 1.00 & 0.58 & 1.00 & 0.54 \\
& 2.2 & 1.00 & 0.56 & 1.00 & 0.57 & 1.00 & 0.54 \\
\midrule
\multirow{4}{*}{\textbf{QCD-E}}
& 1.5 & 1.00 & 0.03 & 1.00 & 0.04 & -- & -- \\
& 1.8 & 1.00 & 0.04 & 1.00 & 0.05 & -- & -- \\
& 2.0 & 1.00 & 0.04 & 1.00 & 0.05 & -- & -- \\
& 2.2 & 1.00 & 0.05 & 1.00 & 0.05 & -- & -- \\
\midrule
\multirow{4}{*}{\textbf{QCD-N}}
& 1.5 & 1.00 & 0.00 & 1.00 & 0.00 & -- & -- \\
& 1.8 & 1.00 & 0.00 & 1.00 & 0.00 & -- & -- \\
& 2.0 & 1.00 & 0.00 & 1.00 & 0.00 & -- & -- \\
& 2.2 & 1.00 & 0.00 & 1.00 & 0.00 & -- & -- \\
\midrule
\multicolumn{8}{l}{\textit{Trimmed (T) with } $\alpha=0.20$} \\
\multirow{4}{*}{\textbf{QCD-T}}
& 1.5 & 1.00 & 0.36 & 1.00 & 0.40 & -- & -- \\
& 1.8 & 1.00 & 0.36 & 1.00 & 0.40 & -- & -- \\
& 2.0 & 1.00 & 0.37 & 1.00 & 0.40 & -- & -- \\
& 2.2 & 1.00 & 0.37 & 1.00 & 0.41 & -- & -- \\
\midrule
\multicolumn{8}{l}{\textit{Trimmed (T) with } $\alpha=0.40$} \\
\multirow{4}{*}{\textbf{QCD-T}}
& 1.5 & 1.00 & 0.54 & 1.00 & 0.56 & -- & -- \\
& 1.8 & 1.00 & 0.53 & 1.00 & 0.56 & -- & -- \\
& 2.0 & 1.00 & 0.53 & 1.00 & 0.56 & -- & -- \\
& 2.2 & 1.00 & 0.55 & 1.00 & 0.56 & -- & -- \\
\bottomrule
\end{tabular*}
\end{table}

\subsection{Simulated EEG data with eye-blink contamination}
EEG signals are often contaminated by artifacts from ocular, muscular, cardiac, and motion related activities. Among these, the electrooculogram (EOG) component (refer to eye-blink artifact hereafter) due to involuntary eyelid movements is the most prevalent. Eye blinks appear as high amplitude, spike-like deflections that contaminate EEG in both time and frequency domains, particularly in the low-frequency range (0--12~Hz) associated with motor control, attention, and drowsiness \citep{chang2016detection}. As they can distort signals up to the alpha band, eye-blink artifacts are a major source of error in band-power analysis. Their removal is therefore essential to improve EEG quality and ensure reliable applications such as driver drowsiness detection, fatigue monitoring, and brain–computer interfaces \citep{maddirala2021eye}. Again, 10 MTS are generate each groups, each MTS is with length between 400 to 2000.

In each group, we fix a contamination proportion $\rho = 0.40$. For each contaminated trial, we draw the number of blinks 
$n_{\mathrm{blink}} \in \{1,2\}$. For each blink, we sample:
\begin{itemize}
  \item a duration $\tau \sim \mathcal{U}(0.20,0.40)\, f_s$, 
        where $\tau$ denotes the blink length in samples 
        (corresponding to 200--400\,ms at sampling rate $f_s$),
  \item a start time $t_0 \in \{1,\dots,T_i-\tau\}$,
  \item a set of frontal channels $\mathcal{I} \subseteq \mathcal{F}$ with 
        $|\mathcal{I}| = \bigl\lceil \rho_{\mathrm{chan}}\, p_{\mathrm{frontal}} \bigr\rceil$,
\end{itemize}
The blink waveform is modeled as a half-sine envelope,
\[
b_q \;=\; \sin\!\Bigl(\tfrac{\pi q}{\tau-1}\Bigr), 
\quad q=0,\dots,\tau-1.
\]

To mimic their large amplitude, we set an amplitude multiplier $\eta \sim \mathcal{U}(4,8)$ and a random polarity $\kappa \in \{-1,+1\}$. For all $q=0,\dots,\tau-1$ and $j \in \mathcal{I}$,
\[
X_{t_0+q,\,j}\ \leftarrow\ X_{t_0+q,\,j}\ +\ \kappa\, \eta\, \hat{\sigma}_j\, b_q.
\]

\begin{figure}[ht]
    \centering
    \includegraphics[width=1\linewidth]{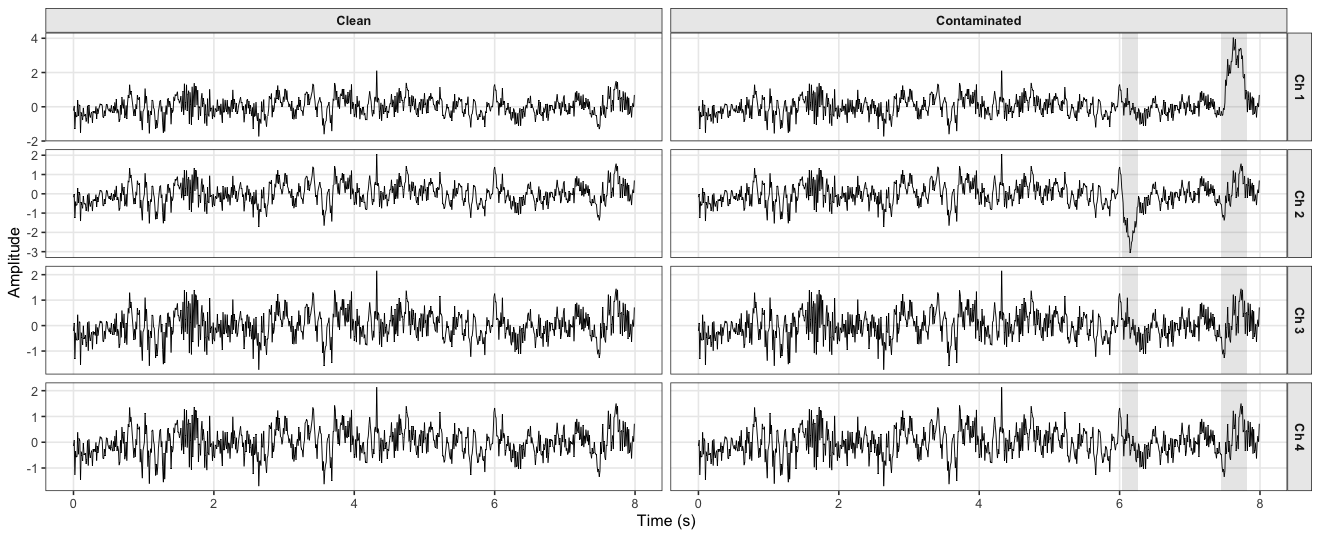}
    \caption{Simulated EEG with eye-blink contamination (one contaminated trial highlighted).}
    \label{eyeblink}
\end{figure}

Thus each selected trial contains one or more transient half-sine deflections localized to frontal electrodes, reproducing the strong low-frequency transients characteristic of real eye-blink artifacts. Fig.~\ref{eyeblink} presents an example of the EEG data with eye-blink contamination.

In Section \ref{burst_contamination_sim}, we already compared against competitors and found that, even with manual tuning, they detected substantially fewer outliers while matching or trailing our methods in clustering accuracy. This section will only focus on FCPCA and RFCPCA and do not repeat the other baselines. Our goal is to test the methods under a different contamination mechanism, and show the improvement of RFCPCA over FCPCA. As shown in Table~\ref{tab:auto_fcpca_rfcpca_tall}, RFCPCA achieves the same clustering accuracy as FCPCA while providing substantially higher outlier detection, highlighting the advantage of robust variants under eye-blink contamination. At the same time, FCPCA also identifies a larger number of outliers in this scenario because the contamination level is as high as 40\%, making outliers easier to detect.

\begin{table}[htbp]
\centering
\caption{FCPCA and RFCPCA with automatic selection of $m$ (and $\alpha$ for trimming). Trial-level contamination $\rho=0.40$. All trials have variable lengths $T\in[400,2000]$.}
\label{auto_fcpca_rfcpca_eyeblink}
\setlength{\tabcolsep}{6pt}
\renewcommand{\arraystretch}{1.08}

\begingroup
\sisetup{round-mode=places,round-precision=2}
\newcolumntype{A}{S[table-format=1.2,round-mode=places,round-precision=2]}
\newcolumntype{O}{S[table-format=1.2,round-mode=places,round-precision=2]}
\newcolumntype{R}{S[table-format=1.2,round-mode=places,round-precision=2]}

\begin{tabular}{l A O @{\,\,\,} R  A O @{\,\,\,} R  A O @{\,\,\,} R}
\toprule
\multirow{2}{*}{Method}
 & \pblock{$p=32$} & \pblock{$p=64$} & \pblock{$p=128$} \\
\cmidrule(lr){2-4}\cmidrule(lr){5-7}\cmidrule(lr){8-10}
 & Acc & Out\% & $\alpha$ & Acc & Out\% & $\alpha$ & Acc & Out\% & $\alpha$ \\
\midrule
FCPCA     & 1.00 & 0.78 & \na & 1.00 & 0.85 & \na & 1.00 & 0.60 & \na \\
RFCPCA\_E & 1.00 & 0.98 & \na & 1.00 & 1.00 & \na & 1.00 & 1.00 & \na \\
RFCPCA\_N & 1.00 & 0.89 & \na & 1.00 & 0.93 & \na & 1.00 & 0.85 & \na \\
RFCPCA\_T & 1.00 & 1.00 & 0.43 & 1.00 & 1.00 & 0.50 & 1.00 & 1.00 & 0.46 \\
\bottomrule
\end{tabular}
\endgroup
\end{table}

From Table \ref{auto_fcpca_rfcpca_eyeblink}, across dimensions, the key difference lies in outlier detection: RFCPCA-E and RFCPCA-T are essentially perfect (Out\% $\approx 0.98$–$1.00$), while RFCPCA-N is high but slightly lower, reflecting the conservativeness of a fixed noise threshold. Plain FCPCA flags only a fraction of contaminated trials (Out\% $=0.60$–$0.85$), consistent with its sensitivity to large low-frequency transients but lack of explicit robustness. The $\alpha$ values selected by RFCPCA-T ($\approx0.43$–$0.50$) adapt closely to the true contamination level $\rho=0.40$, with small deviations attributable to random blink count, timing, and channel selection. Overall, the robust variants reliably isolate eye-blink contamination under variable $T_i$ without manual tuning, which is the intended operating regime for real EEG.

\section{Exploring the mapping between EEG and alertness in a driving experiment}\label{application}

In this section, we consider a publicly available EEG data\footnote[1]{\url{https://figshare.com/articles/dataset/EEG_driver_drowsiness_dataset/14273687?file=30707285}}. The EEG Driver Drowsiness Dataset contains multi-channel EEG recordings from 11 subjects performing a sustained-attention driving task, with labels indicating alert and drowsy states. The data include 2022 3-second EEG samples (128 Hz, 30 channels). We evaluate clustering accuracy  using three metrics: the RI, the adjusted Rand index (ARI), and the fuzzy Rand index (RIF). The same dataset is used here as \citet{ma2025fcpca} to directly show the gains in robustness.

\subsection{The clustering performance on all subjects}
Table \ref{real_app_AD} shows that all three robust extensions generally outperform FCPCA across subjects. We now focus on Subjects 6 and 9, which exhibit the most pronounced gains in accuracy under the RFCPCA variants.

\begin{table}[htbp]
\centering
\small
\caption{Comparison of clustering accuracy (RI / ARI / RIF) per subject.
Bold indicates an RFCPCA variant strictly better than FCPCA on all three metrics.}
\label{real_app_AD}
\setlength{\tabcolsep}{6pt}           
\renewcommand{\arraystretch}{1.25}    
\begin{tabular}{
  c c
  S[table-format=1.2] S[table-format=1.2] S[table-format=1.2]
  S[table-format=1.2] S[table-format=1.2] S[table-format=1.2]
  S[table-format=1.2] S[table-format=1.2] S[table-format=1.2]
  S[table-format=1.2] S[table-format=1.2] S[table-format=1.2]
}
\toprule
\multicolumn{2}{c}{} &
\multicolumn{3}{c}{\textbf{FCPCA}} &
\multicolumn{3}{c}{RFCPCA-E} &
\multicolumn{3}{c}{RFCPCA-T} &
\multicolumn{3}{c}{RFCPCA-N} \\
\cmidrule(lr){3-5}\cmidrule(lr){6-8}\cmidrule(lr){9-11}\cmidrule(lr){12-14}
Subject \# & Size &
{RI} & {ARI} & {RIF} &
{RI} & {ARI} & {RIF} &
{RI} & {ARI} & {RIF} &
{RI} & {ARI} & {RIF} \\
\midrule
1  & 188 & 0.740 & 0.240 & 0.510 & \bfseries 0.820 & \bfseries 0.390 & \bfseries 0.700 & \bfseries 0.780 & \bfseries 0.320 & \bfseries 0.550 & \bfseries 0.810 & \bfseries 0.350 & \bfseries 0.640 \\
2  & 132 & 0.890 & 0.620 & 0.820 & 0.890 & 0.600 & 0.810 & \bfseries 0.910 & \bfseries 0.660 & \bfseries 0.890 & 0.890 & 0.610 & 0.820 \\
3  & 150 & 0.550 & 0.010 & 0.890 & 0.550 & 0.010 & 0.900 & \bfseries 0.690 & \bfseries 0.140 & \bfseries 0.950 & \bfseries 0.630 & \bfseries 0.060 & \bfseries 0.910 \\
4  & 148 & 0.570 & 0.020 & 0.580 & \bfseries 0.600 & \bfseries 0.040 & \bfseries 0.590 & \bfseries 0.690 & \bfseries 0.130 & \bfseries 0.650 & \bfseries 0.700 & \bfseries 0.160 & \bfseries 0.660 \\
5  & 224 & 0.880 & 0.560 & 0.740 & 0.880 & 0.570 & 0.880 & 0.880 & 0.570 & 0.720 & 0.880 & 0.560 & 0.730 \\
6  & 166 & 0.510 & 0.000 & 0.520 & \bfseries 0.840 & \bfseries 0.450 & \bfseries 0.940 & \bfseries 0.990 & \bfseries 0.880 & \bfseries 0.980 & \bfseries 0.710 & \bfseries 0.170 & \bfseries 0.640 \\
7  & 102 & 0.580 & 0.020 & 0.660 & \bfseries 0.640 & \bfseries 0.090 & \bfseries 0.690 & \bfseries 0.590 & \bfseries 0.040 & \bfseries 0.670 & 0.580 & 0.010 & 0.650 \\
8  & 264 & 0.660 & 0.090 & 0.610 & \bfseries 0.690 & \bfseries 0.120 & \bfseries 0.680 & \bfseries 0.690 & \bfseries 0.140 & \bfseries 0.680 & 0.650 & 0.080 & 0.600 \\
9  & 314 & 0.750 & 0.260 & 0.840 & \bfseries 0.880 & \bfseries 0.550 & \bfseries 0.850 & \bfseries 0.920 & \bfseries 0.680 & \bfseries 0.850 & \bfseries 0.920 & \bfseries 0.690 & \bfseries 0.890 \\
10 & 108 & 1.000 & 1.000 & 0.920 & 1.000 & 1.000 & 0.930 & 0.990 & 0.960 & 0.940 & 1.000 & 1.000 & 0.940 \\
11 & 226 & 0.930 & 0.750 & 0.860 & 0.930 & 0.720 & 0.950 & \bfseries 0.960 & \bfseries 0.850 & \bfseries 0.970 & 0.930 & 0.710 & 0.870 \\
\midrule
Mean &     & 0.730 & 0.320 & 0.720 & \bfseries 0.790 & \bfseries 0.410 & \bfseries 0.820 & \bfseries 0.830 & \bfseries 0.490 & \bfseries 0.800 & \bfseries 0.790 & \bfseries 0.400 & \bfseries 0.760 \\
\bottomrule
\end{tabular}
\end{table}

\subsection{Analysis of Subject 6}
\begin{figure}[htbp]
    \centering
    \includegraphics[width=1\linewidth]{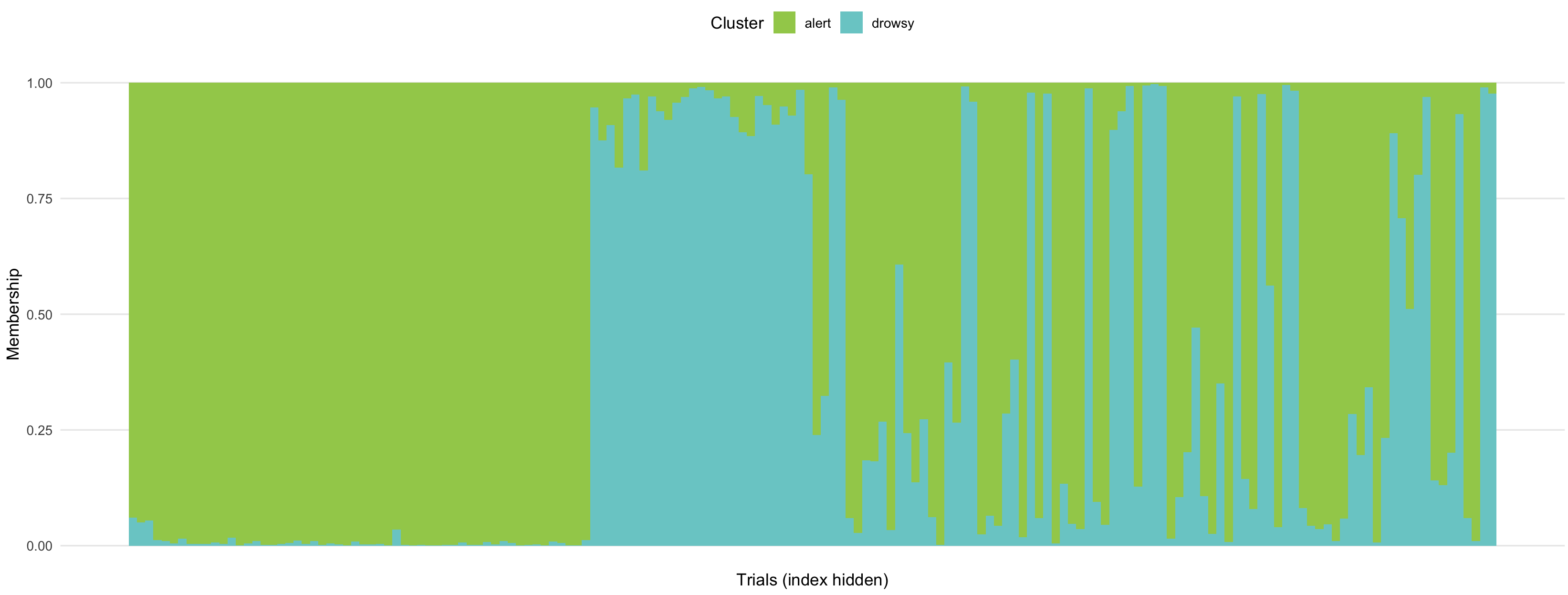}
    \caption{Membership matrix of Subj 6 using FCPCA}
    \label{fcpca_mem_6}
\end{figure}

The ground-truth partition is simple: the first half of trials are alert and the second half are drowsy. However, the FCPCA membership matrix in Fig.~\ref{fcpca_mem_6} does not recover this step change: memberships drift across the sequence, so no clear block forms at the midpoint. A plausible reason is that the quadratic loss in FCPCA is dominated by many borderline/transitional windows and within-state heterogeneity. Suppose physiology shifts gradually while the labels switch at a hard boundary, numerous windows straddle the change and appear ambiguous. In that case, fitting them under squared loss pulls the two CPCA subspaces toward a common average, reducing separation. Consequently, the memberships in Fig.~\ref{fcpca_mem_6} are low-contrast and unstable.

\begin{figure}[htbp]
    \centering
    \includegraphics[width=1\linewidth]{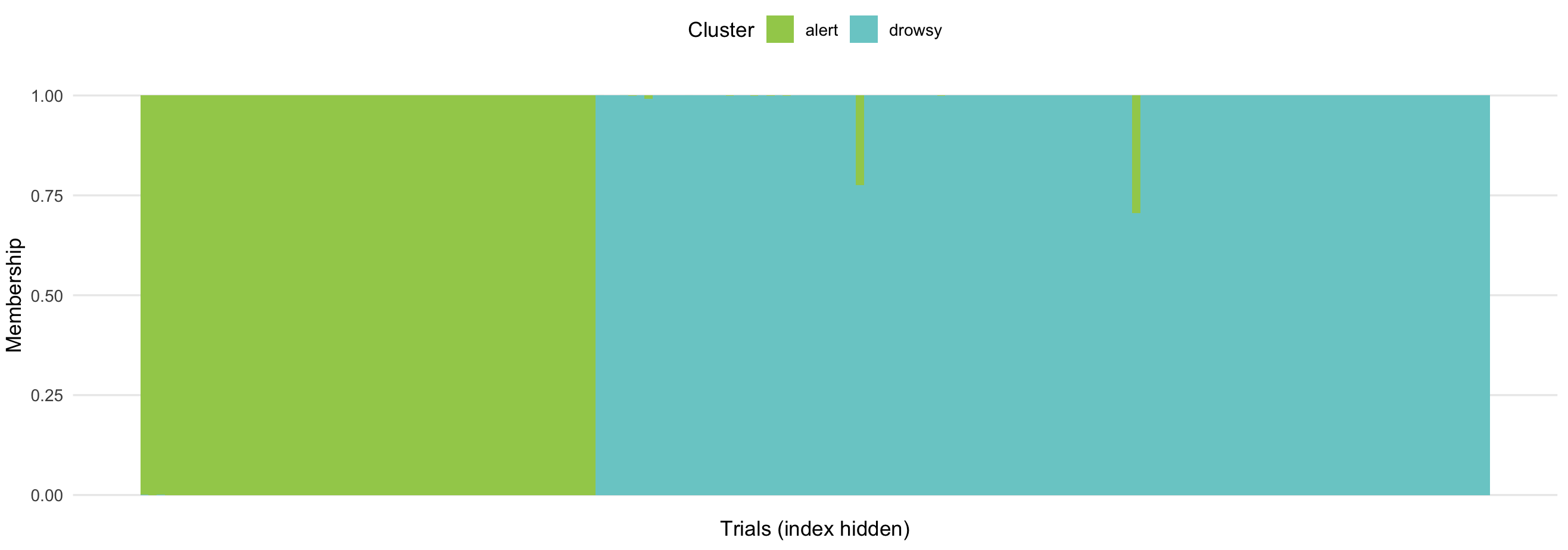}
    \caption{Membership matrix of Subj 6 using RFCPCA-E}
    \label{rfcpca_e_mem_6}
\end{figure}

Moving to RFCPCA-E, Fig.~\ref{rfcpca_e_mem_6} shows that the exponential loss down-weights poorly fitted windows, so the subspaces are estimated mainly from the core trials. The result suggests a crisp, step-like partition with a clear split near the midpoint. Some short mistakes remain in the middle, a brief stretch that should be “alert” is assigned to “drowsy.” These misassignments are consistent with lingering transitional/heterogeneous windows: they are not gross outliers (as corroborated by the dispersed, low-noise mass under RFCPCA-N later), but still sit closer to the drowsy subspace after reweighting, so they flip labels locally, even though the global partition is much cleaner.
 
\begin{figure}[htbp]
    \centering
    \includegraphics[width=1\linewidth]{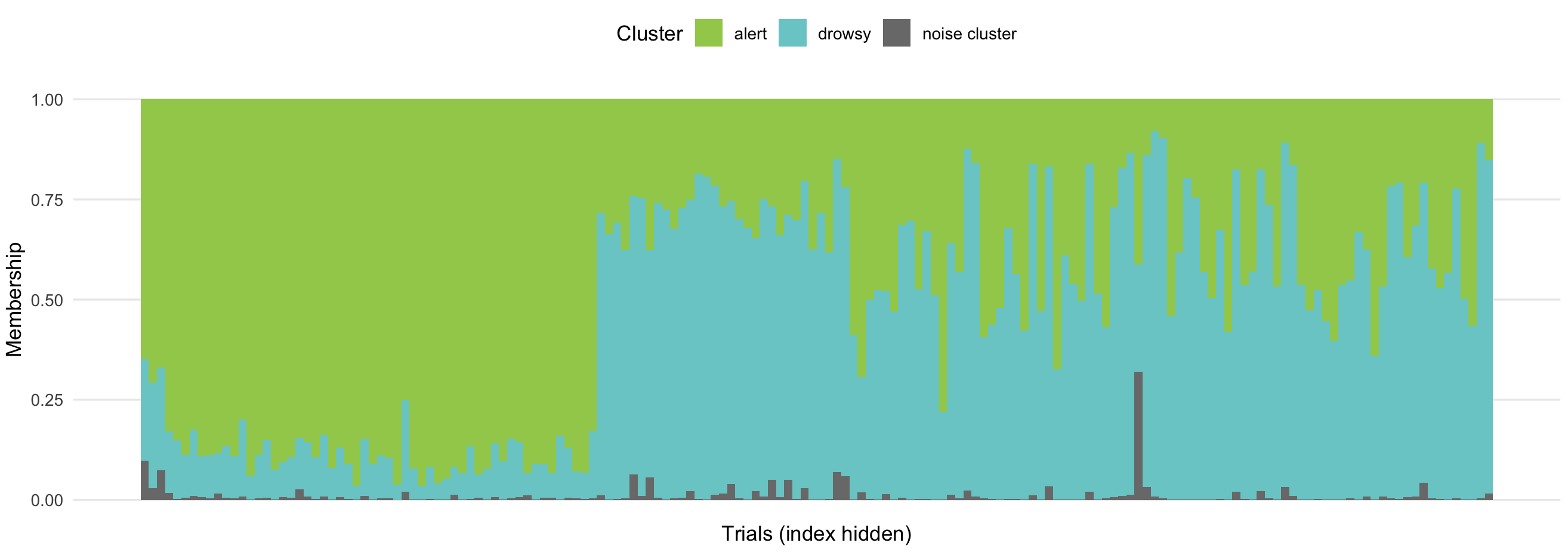}
    \caption{Membership matrix of Subj 6 using RFCPCA-N}
    \label{rfcpca_n_mem_6}
\end{figure}

With RFCPCA\text{-}N (Fig.~\ref{rfcpca_n_mem_6}), there is no obvious outlier trail.
Noise memberships are small and dispersed. Accuracy improves because trials that assign some mass to the noise cluster receive reduced effective weight when
estimating the projection axes, so ambiguous or slightly outlying trials contribute
less to the subspace estimation, yielding cleaner projection axes.

\begin{figure}[htbp]
    \centering
    \includegraphics[width=1\linewidth]{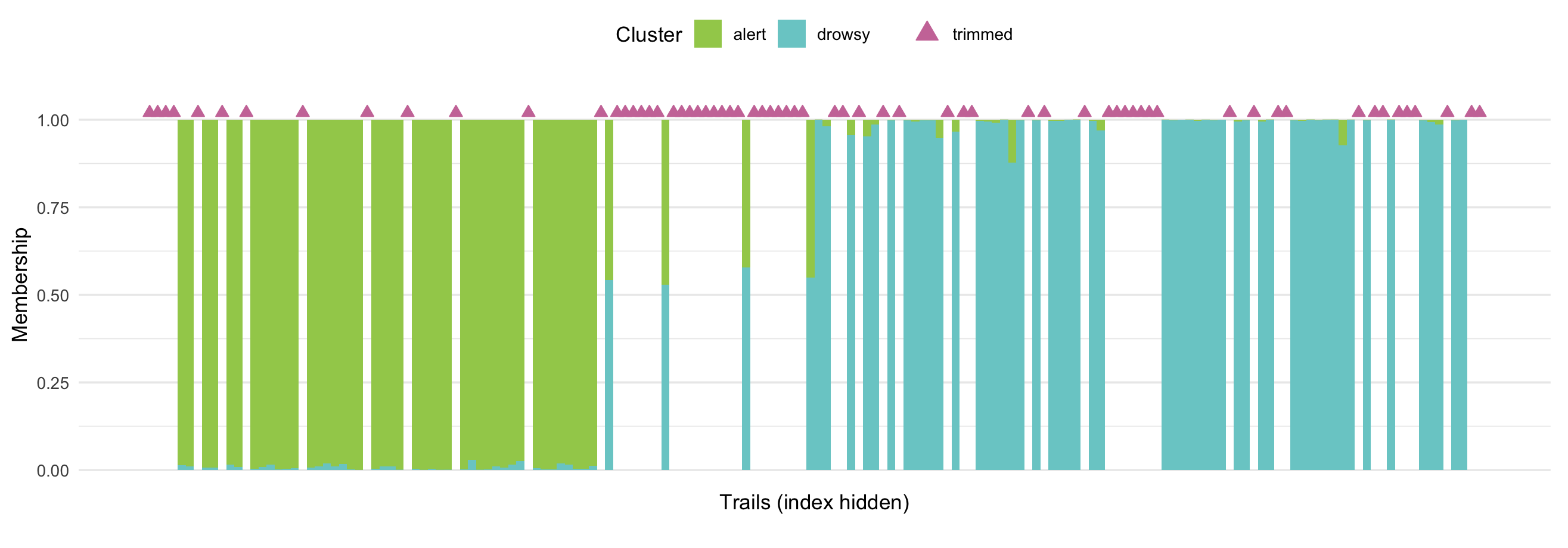}
    \caption{Membership matrix of Subj 6 using RFCPCA-T}
    \label{rfcpca_t_mem_6}
\end{figure}

As shown in Fig.~\ref{rfcpca_t_mem_6}, the trimmed windows (triangle markers) are concentrated in the middle portion of the session, precisely where
RFCPCA-E  and RFCPCA-N  still make mistakes by assigning some “alert” trials to “drowsy.” By excluding these trials from the subspace estimation, RFCPCA-T achieves the highest clustering accuracy among the methods (near perfect). Notably, the same trial that carries a high noise membership under RFCPCA-N is included in the trimmed set of RFCPCA-T, indicating that the two robust strategies are aligned to identify difficult windows and that trimming resolves the ambiguity.

\subsection{Analysis of Subject 9}
\begin{figure}[htbp]
    \centering
    \includegraphics[width=1\linewidth]{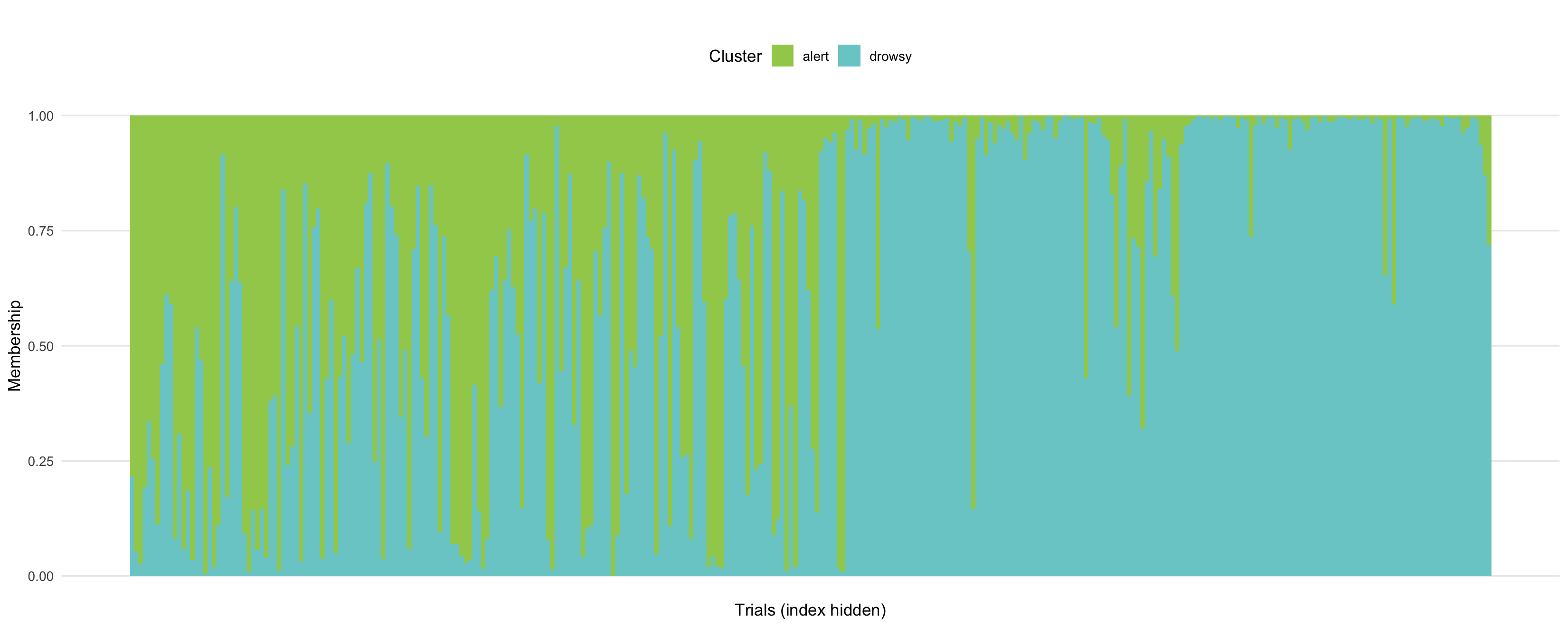}
    \caption{Membership matrix of Subj 9 using FCPCA}
    \label{fcpca_mem_9}
\end{figure}

The membership matrix in Fig.~\ref{fcpca_mem_9} shows the membership matrix using FCPCA.   Many windows carry mixed memberships, and the transition region is diffuse, indicating that the two CPCA subspaces are partially averaged when confronted with gradual state drift and within-state heterogeneity.

\begin{figure}[htbp]
    \centering
    \includegraphics[width=1\linewidth]{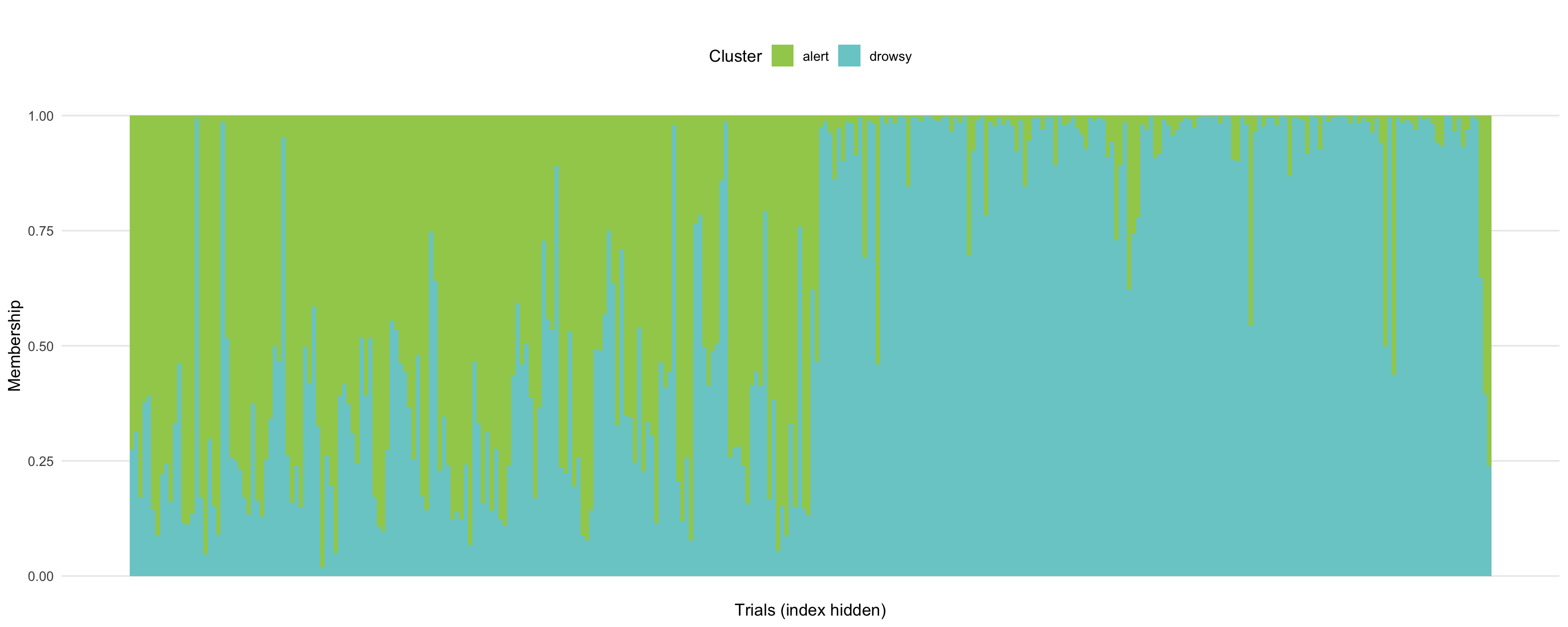}
    \caption{Membership matrix of Subj 9 using RFCPCA-E}
    \label{rfcpca_e_mem_9}
\end{figure}

With RFCPCA-E (Fig.~\ref{rfcpca_e_mem_9}), down–weighting poorly fitted trails improves both brain states segment: most of the first part is now cleanly assigned to alter rather than being intermittently misclassified, and the second half forms a much clearer drowsy block with only occasional, narrow spikes. The transition between the two segments becomes markedly more step-like, indicating better separation of the subspaces.

\begin{figure}[htbp]
    \centering
    \includegraphics[width=1\linewidth]{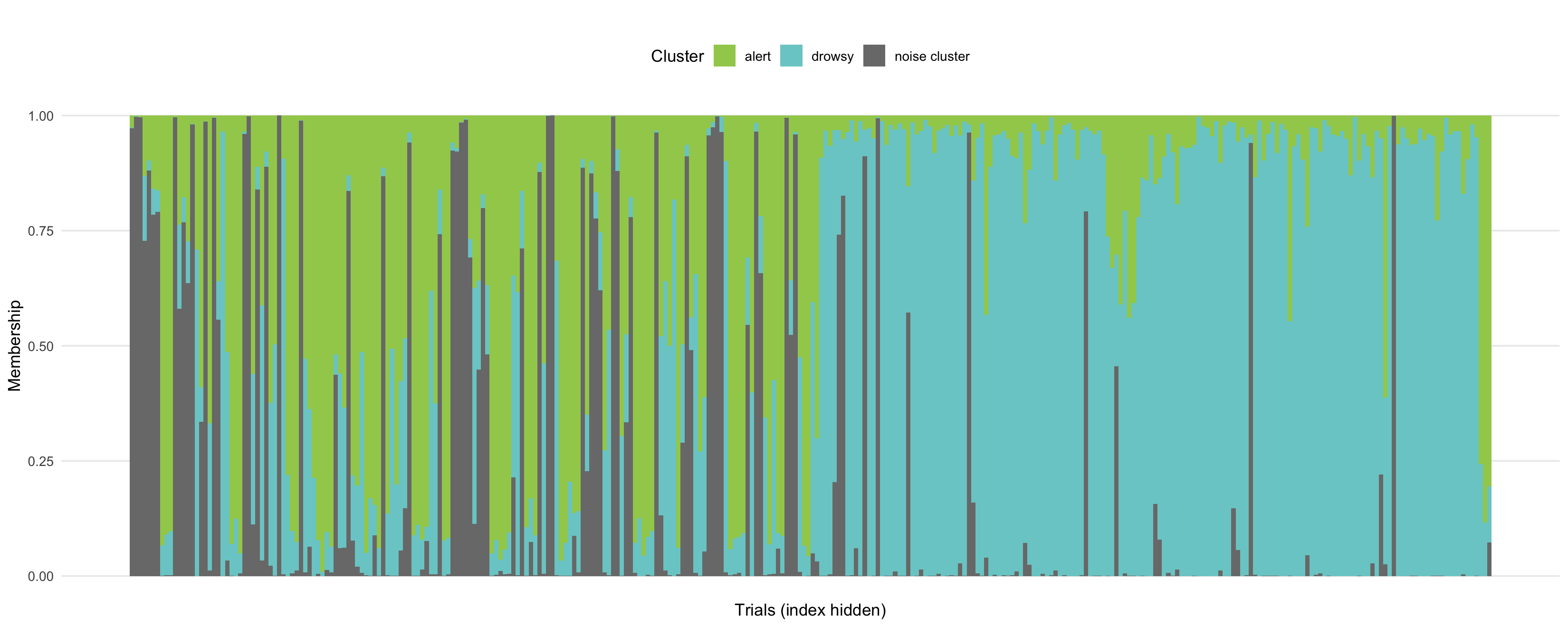}
    \caption{Membership matrix of Subj 9 using RFCPCA-N}
    \label{rfcpca_n_mem_9}
\end{figure}

In Fig.~\ref{rfcpca_n_mem_9}, trials flagged as outliers are frequent, especially in
the first third. Several trials exhibit almost-crisp membership in the noise
cluster, indicating activity that is highly inconsistent with both the alert and
drowsy subspaces, either strong artifacts or an unmodeled brain state not captured
by the two-class setup—and thus potentially meaningful for downstream analysis.

\begin{figure}[htbp]
    \centering
    \includegraphics[width=1\linewidth]{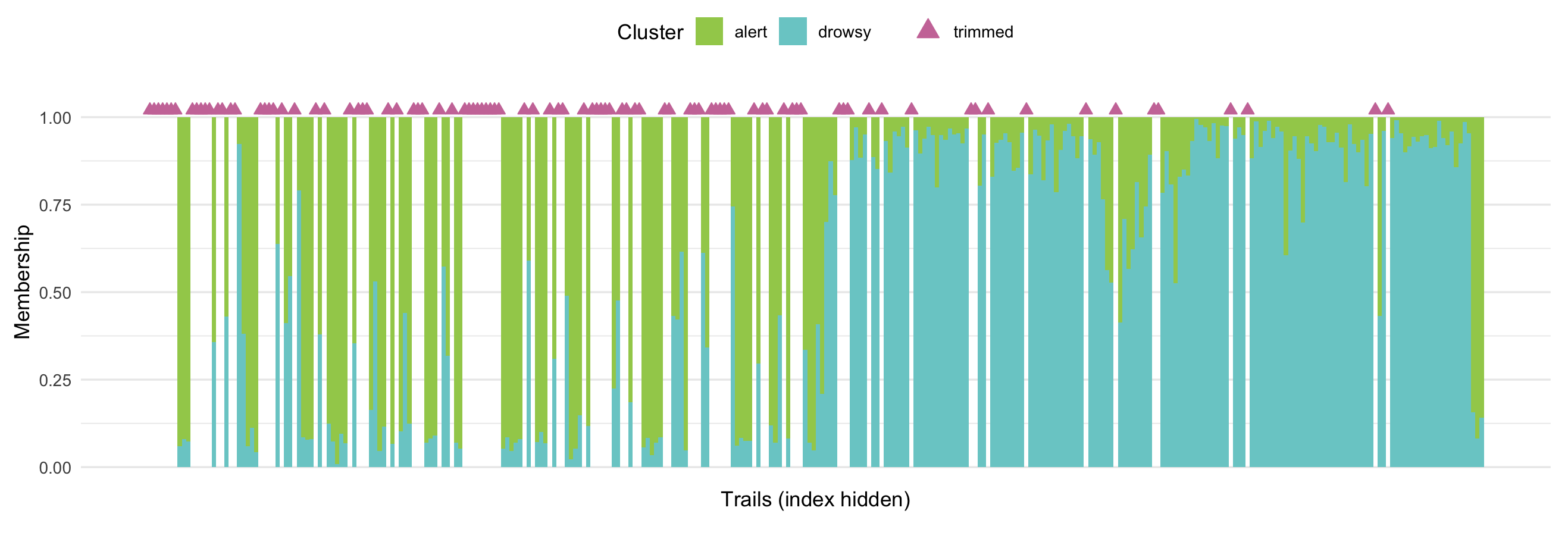}
    \caption{Membership matrix of Subj 9 using RFCPCA-T}
    \label{rfcpca_t_mem_9}
\end{figure}

Lastly, in Fig.~\ref{rfcpca_t_mem_9}, the trimmed windows (triangle markers) concentrate in the early, more variable portion and around the transition. Trimming removes these discordant windows from the subspace estimation and produces the cleanest two-block structure and the best accuracy. Quantitatively, RFCPCA-T trims about 30\% of trials, whereas RFCPCA-N flags about 23\% as noise. Interesting, all RFCPCA-N outliers are included in the RFCPCA-T trimmed set (100\% overlap), while the remaining ~7\% trimmed windows are additional borderline cases that RFCPCA-N treats with small, dispersed noise mass. This alignment explains why both methods identify the same difficult region, and why trimming resolves the residual ambiguity.

\subsection{Discriminating subspaces and further insights with RFCPCA-N on Subject 9}
Subject~9 contains a sizable fraction of trials with non-negligible noise membership under RFCPCA-N. Our goals here are twofold: (i) verify that the noise component corresponds to dynamics that are genuinely different from the normal physiological states; and (ii) measure the extent of overlap (shared axes) between the alert and drowsy subspaces and verify that they remain separable, as expected under fuzzy, graded transitions.

\begin{figure}
    \centering
    \includegraphics[width=1\linewidth]{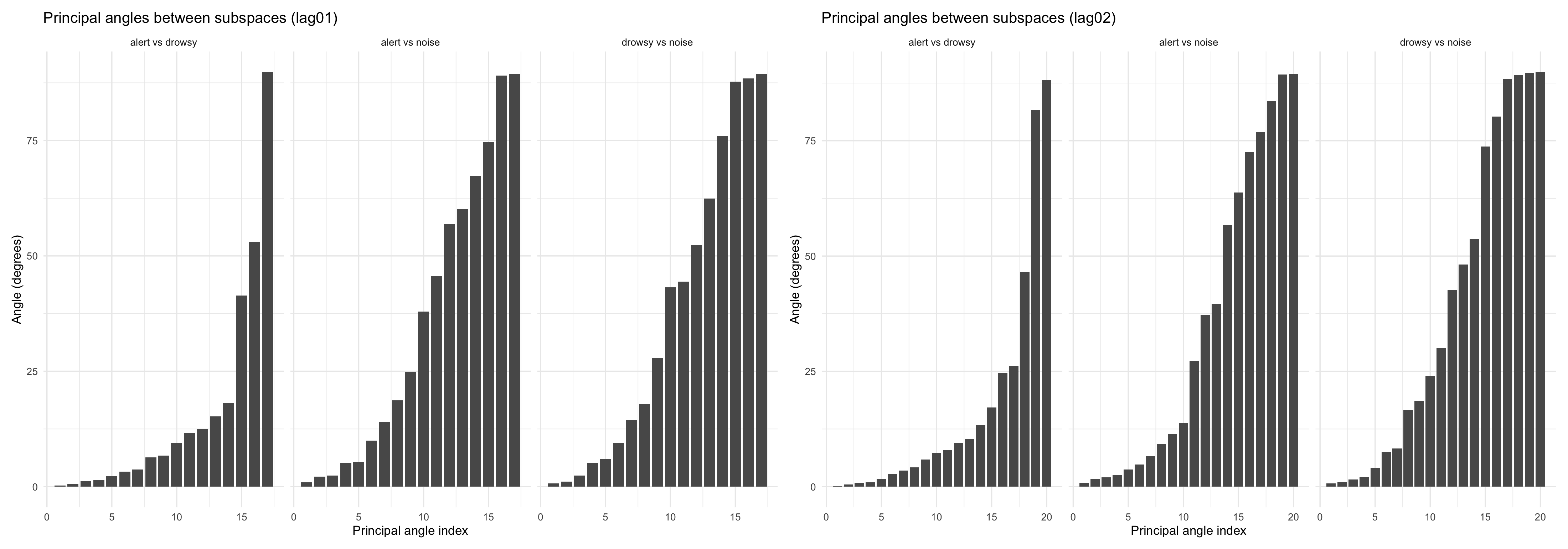}
    \caption{Principal angles (in increasing order) between alert, drowsy, and noise subspaces at lag 0–1 and lag 0–2}
    \label{principal_angles}
\end{figure}

To quantify the degree of similarity and separation among the estimated
subspaces, we first computed the principal angles between pairs of subspaces
(alert, drowsy, and noise). Suppose $\mathcal{B}_a$ and
$\mathcal{B}_b$ denote two sets of orthonormal basis vectors obtained from the
projection axes of different clusters. The cosines of the principal angles are
then given by the singular values of $\mathcal{B}_a^{\top}\mathcal{B}_b$. Small angles
reveal overlapping directions (shared latent dynamics), while large angles
indicate nearly orthogonal directions (distinct dynamics). Fig.~\ref{principal_angles}
demonstrates that the {alert} and {drowsy} subspaces share several
common axes, showing partially overlapping activity. whereas the {noise} subspace is much more different from both, indicating that the noise subspace captures dynamics distinct from both brain states. The separation is more pronounced at lag~0--2, suggesting that temporal dependencies enhance discrimination.

We then examined the contribution of individual EEG channels to the formation
of each subspace. For channel $j$ in cluster $s$, the contribution is defined as
$$\text{contrib}_{j,s} = \sum_{k} C_s(j,k)^2,$$
where $C_s(j,k)$ denotes the loading of channel $j$ on the $k$-th retained basis
vector of the subspace associated with cluster $s$. This measure reflects how strongly each electrode shapes the discriminating axes. 

\begin{figure}[htbp]
    \centering
    \includegraphics[width=1\linewidth]{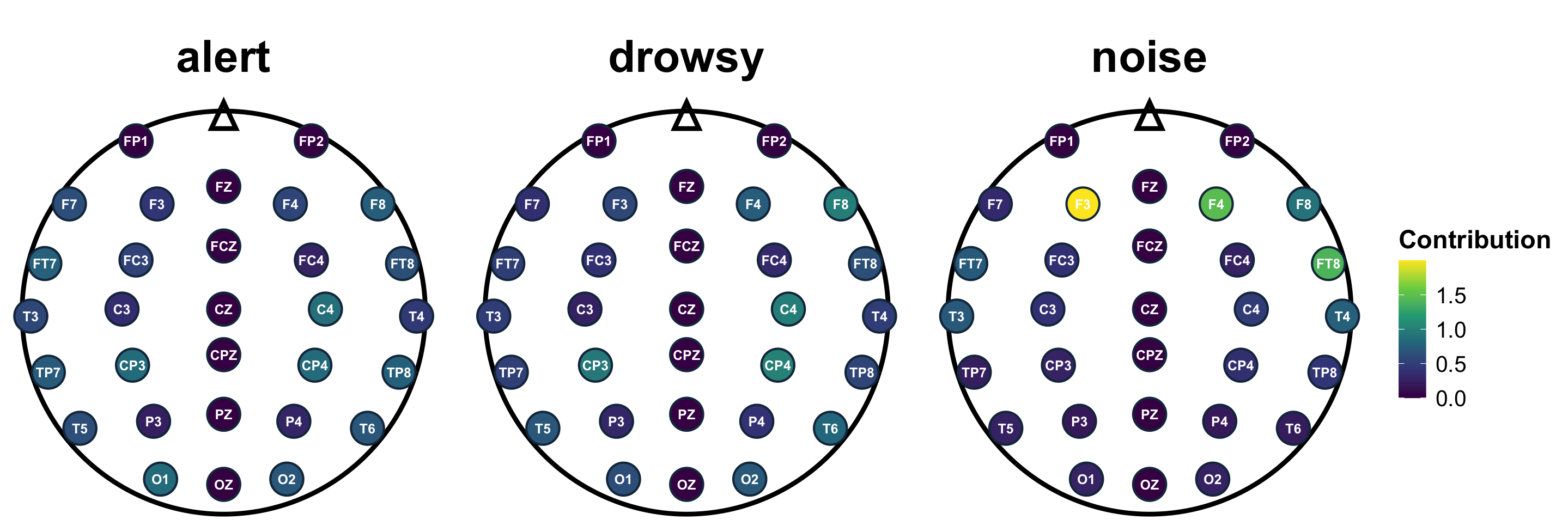}
    \caption{EEG Channel Contributions to alert, drowsy, and noise subspaces using lag 0–1}
    \label{channel1}
\end{figure}

\begin{figure}[htbp]
    \centering
    \includegraphics[width=1\linewidth]{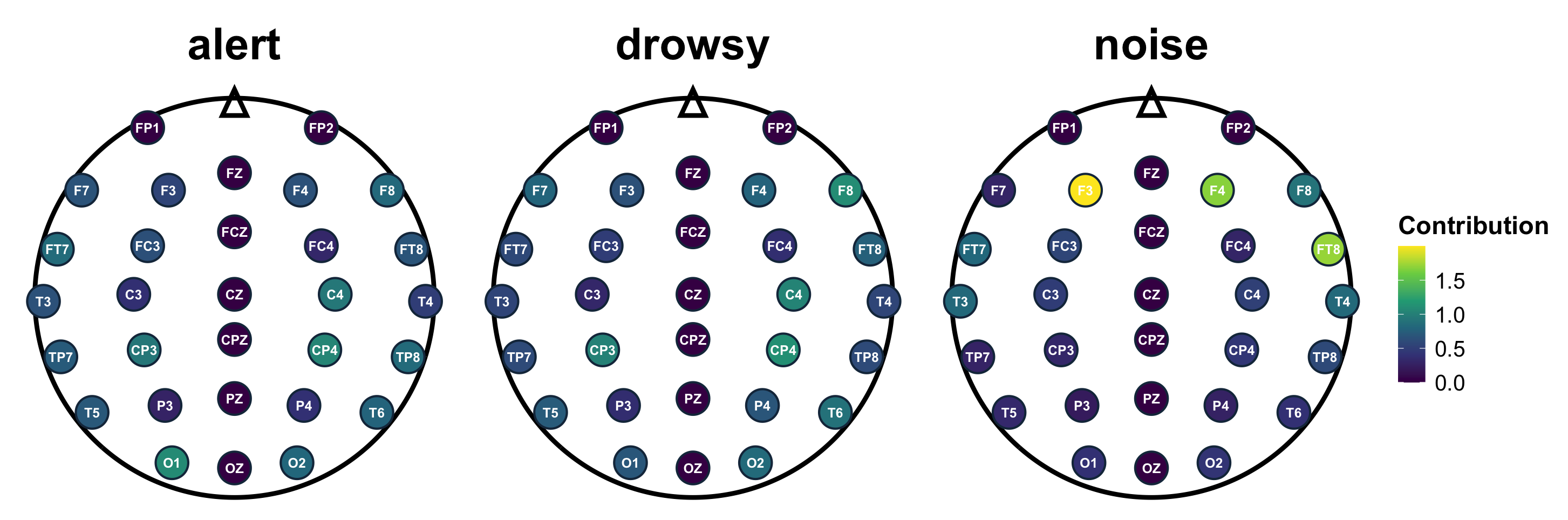}
    \caption{EEG Channel Contributions to alert, drowsy, and noise subspaces using lag 0–2}
    \label{channel2}
\end{figure}

Fig.~\ref{channel1} and \ref{channel2} display the contributions of all 30 EEG channels to the alert, drowsy, and noise subspaces using lag 0–1 and lag 0–2, respectively. The alert and drowsy maps share broadly similar patterns, further verifying separable but overlapping patterns consistent with transitional or mixed states. Despite this similarity, the two states also show differences in discrimination (e.g., FT7, F8, O1).

The noise subspace remains distinct across both lags, with higher weights concentrated at frontal/temporal sites (e.g., F3, F4, FT8), suggesting that it primarily captures artifacts or other unsystematic activity rather than task-related dynamics.

\subsection{Summary of the EEG data: robust gains and fuzzy insights}
Although the dataset is preprocessed and should not contain obvious outliers, RFCPCA still yields clear gains in clustering accuracy. This indicates that robustness is beneficial not only under heavy contamination but also for subtle irregularities and residual noise that persist after preprocessing.

Beyond accuracy, the fuzzy formulation adds interpretability. First, fuzzy memberships capture ambiguous or transitional states between clusters—particularly relevant because brain activity seldom switches abruptly between alertness and drowsiness. Second, membership magnitudes quantify assignment uncertainty, highlighting series that, while not gross outliers, deviate from dominant patterns. Third, secondary memberships reveal overlap and shared structure across clusters, yielding a richer description than crisp partitions. Finally, the robust extensions detect partially outlying series along a continuum rather than forcing a binary inlier/outlier decision.

\section{Conclusion}\label{conclusion}

We presented a robust fuzzy subspace clustering framework for MTS that builds on FCPCA and introduces three complementary variants, RFCPCA-E (robust metric), RFCPCA-N (noise cluster), and RFCPCA-T (trimming), all with automatic selection of the number of clusters $S$, the fuzziness parameter $m$, and, when applicable, the trimming fraction $\alpha$. The approach summarizes each object with covariance features, accommodates variable lengths $T_i$, and uses fuzzy memberships to quantify assignment uncertainty and handle atypical trials.

Across synthetic EEG-like settings with artifacts and in real EEG, the robust variants improve clustering accuracy relative to FCPCA while providing principled outlier handling. In Simulation~1 (burst contamination), RFCPCA consistently retained the clean structure and isolated contaminated trials. In Simulation~2 (eye-blink contamination), which stresses low-frequency, trial-level transients with varying lengths, RFCPCA again achieved perfect or near-perfect accuracy and markedly higher outlier detection than FCPCA under fully automatic tuning. On the real dataset, robustness remained beneficial even after standard preprocessing, and the fuzzy memberships improved interpretation by indicating gradual transitions between clusters; additionally, the estimated noise subspace was distinct from physiologic subspaces, supporting a meaningful separation between structured signal and contamination.

Despite these strengths, the framework is primarily second-order: covariance summaries emphasize amplitude cofluctuations and may miss discriminative phase or cross-lag structure when such information is essential. It also presumes a stationary core process under moderate contamination; performance can degrade when artifacts dominate many trials or persist for long durations. In addition, eigendecomposition of high-dimensional covariances becomes costly as $p$ and the number of trials grow. Future work includes improving scalability via shrinkage covariances, randomized eigensolvers, and GPU acceleration, and enhancing robustness to missing or noisy channels. Moreover, the robust fuzzy clustering-based MTS forecasting algorithm can also be considered.

\section*{Acknowledgments}
This research was supported by King Abdullah University of Science and Technology (KAUST). 


\section*{Conflict of interest statement}
The authors declare that they have no known competing financial
interests or personal relationships that could have appeared to
influence the work reported in this paper.


\section*{Code availability}
All three RFCPCA methods are coded in R and are freely available on the following GitHub repository: \url{https://github.com/arbitraryma/RFCPCA.git}. For any questions or inquiries, please contact the corresponding author.

\pagebreak  
\newpage
\addcontentsline{toc}{chapter}{Bibliography}

\bibliographystyle{biom}
\bibliography{paper}

\end{document}